\documentclass[aps,prd,twocolumn,groupedaddress,reprint,draft,noeprint,longbibliography,superscriptaddress,nofootinbib,balancelastpage]{revtex4-2}

\usepackage[hidelinks,draft=false]{hyperref}
\usepackage{color}
\usepackage[normalem]{ulem}
\usepackage{enumitem}
\usepackage{verbatim}
\usepackage{afterpage}

\usepackage{amsmath, amssymb}
\usepackage{mathtools}
\usepackage{empheq}
\renewcommand{\bm}[1]{{\mbox{\boldmath $ #1 $}}}
\allowdisplaybreaks
\usepackage{graphicx}
\setkeys{Gin}{draft=false}

\usepackage{braket}

\makeatletter
\renewcommand{\env@cases}[1][@{}l@{\quad}l@{}]{%
	\let\@ifnextchar\new@ifnextchar
	\left\lbrace
	\def\arraystretch{1.2}%
	\array{#1}%
}
\usepackage{tensor}
\newcommand{\idx}{\indices}
\newcommand{\de}{\partial}
\newcommand{\De}{\mathcal{D}}
\newcommand{\di}[1][]{\mathrm{d}^{#1}}
\newcommand\numberthis{\addtocounter{equation}{1}\tag{\theequation}}
\makeatletter
\newsavebox{\supsub@sup}
\newsavebox{\supsub@sub}
\newlength{\supsub@wd}

\newcommand{\supsub}[2]{%
	\sbox\supsub@sup{$\m@th\scriptstyle#1$}%
	\sbox\supsub@sub{$\m@th\scriptstyle#2$}%
	\setlength{\supsub@wd}{\wd\supsub@sup}%
	\ifdim\supsub@wd<\wd\supsub@sub
	\setlength{\supsub@wd}{\wd\supsub@sub}%
	\fi
	^{\makebox[\supsub@wd]{\usebox{\supsub@sup}}}%
	_{\makebox[\supsub@wd]{\usebox{\supsub@sub}}}%
}
\makeatother
\newcommand{\Christoffel}[2]{\left\{\text{}\supsub{#1}{#2}\right\}}

\usepackage[utf8]{inputenc}
\usepackage{amsthm}

\usepackage{array}
\usepackage{hhline}
\usepackage{tabularx}
\newcolumntype{Y}{>{\centering\arraybackslash}X}
\newcolumntype{H}{@{}>{\lrbox0}l<{\endlrbox}}
\usepackage{longtable}
\usepackage{makecell}
\usepackage{placeins} 
\def\LT@LR@e{\LTleft\z@   \LTright\z@}%

\usepackage[usestackEOL]{stackengine} 

\usepackage{cleveref} 
\crefname{appch}{Appendix}{Appendices}

\newcounter{magicrownumbers}
\newcommand{\rownumber}{\refstepcounter{magicrownumbers}\arabic{magicrownumbers}}

\usepackage{ifdraft}

\ifdraft{
	\usepackage[colorinlistoftodos]{todonotes}
	\newcommand{\mynote}[2][false]{
		\begingroup
		\ifthenelse{\equal{#1}{false}}{\def\clr{green}}{\def\clr{red}}
		\todo[author=ycl54,size=\small,inline,color=\clr!40]{#2}
		\endgroup
	}
	\newcommand{\annote}[2][false]{
		\begingroup
		\ifthenelse{\equal{#1}{false}}{\def\clr{green}}{\def\clr{red}}
		\todo[author=anlas,size=\small,inline,color=\clr!60]{#2}
		\endgroup
	}
}{
	\newcommand{\mynote}[2][false]{}
	\newcommand{\annote}[2][false]{}

}

\newcommand{\numCriticalNotRootNotZero}{862}
\newcommand{\numNoGhostNoTachyon}{168}
\newcommand{\numNoGhostNoTachyonPCR}{40}
\newcommand{\numNoGhostNoTachyonPCRMlMv}{6}
\newcommand{\numNoGhostNoTachyonPCRMl}{11}
\newcommand{\numNoGhostNoTachyonPCRMv}{23}

\def\matri#1{{\mbox{\sf  #1}}}

\newcommand{\myDel}[1]{{\color{red}\ifmmode\text{\sout{$#1$}}\else\sout{#1}\fi}}

\newcommand{\mphDel}[1]{{\color{red}\ifmmode\text{\sout{$#1$}}\else\sout{#1}\fi}}

\begin{document}


\title{Ghost and tachyon free Weyl gauge theories: a systematic approach}


\author{Yun-Cherng~\surname{Lin}}
\email{ycl54@mrao.cam.ac.uk}
\affiliation{Astrophysics Group, Cavendish Laboratory, JJ Thomson Avenue,
Cambridge CB3 0HE, UK}
\affiliation{Kavli Institute for Cosmology, Madingley Road, Cambridge CB3 0HA, UK}
\author{Michael P.~\surname{Hobson}}
\email{mph@mrao.cam.ac.uk}
\affiliation{Astrophysics Group, Cavendish Laboratory, JJ Thomson Avenue,
Cambridge CB3 0HE, UK}
\author{Anthony N.~\surname{Lasenby}}
\email{a.n.lasenby@mrao.cam.ac.uk}
\affiliation{Astrophysics Group, Cavendish Laboratory, JJ Thomson Avenue,
Cambridge CB3 0HE, UK}
\affiliation{Kavli Institute for Cosmology, Madingley Road, Cambridge 
CB3 0HA, UK}


\date{\today}

\begin{abstract}
We investigate the particle content of parity-preserving
Weyl gauge theories of gravity (WGT$^+$) about a
Minkowski background. Within a subset of the full theory, we use a
systematic method previously presented in \cite{Lin2019a} to determine
\numCriticalNotRootNotZero{} critical cases for
which the parameter values in the action lead to additional gauge
invariances.  We find that \numNoGhostNoTachyon{}
of these cases are free of ghosts and tachyons, provided the
parameters satisfy certain conditions that we also determine. We
further identify \numNoGhostNoTachyonPCR{} of these cases that are
also propagating power-counting renormalizable and determine the corresponding
conditions on the parameters. Of these theories,
\numNoGhostNoTachyonPCRMl{} have only massless tordion propagating
particles, \numNoGhostNoTachyonPCRMv{} have only a massive tordion
propagating mode, and \numNoGhostNoTachyonPCRMlMv{} have both. We also
repeat our analysis for WGT$^+$ with vanishing torsion or curvature,
respectively. We compare our findings with the very few previous
results in the literature.
\end{abstract}

\pacs{Add PACS}

\maketitle


\section{Introduction}

In recent papers \cite{Lin2019a,Lin2020}, we presented a systematic
method for identifying the ghost-and-tachyon-free critical cases of
parity-preserving gauge theories of gravity, and applied it to
parity-preserving Poincar\'e gauge theory (PGT$^+$). We found 450
critical cases (which possess additional gauge invariances) that are
free of ghosts and tachyons. We also considered the superficial
  renormalizability by power counting of a subset of these
  unitary theories for which there are no terms in the gauge-fixed
  Lagrangian that mix different fields. While not stated
  explicitly in \cite{Lin2020}, 4 of the theories in that paper
  (cases 9, 10, 11 and 13, which have only massless modes) satisfy the
  original criterion used by Sezgin in \cite{Sezgin1980} to be power
  counting renormalizable (PCR). Moreover, we found a further 54
  theories that satisfy a less restrictive criterion, which in
  addition permits the presence of modes that are non-propagating at
  large momenta (for which the propagator decays no faster than $\sim
  k^0$), since these should then completely decouple from the rest of
  the theory; this is termed `the alternative PCR criterion' in
  \cite{Lin2020}, but here (and henceforth) we shall instead refer to as
  `propagating power counting renormalizable' (PPCR) to avoid
  confusion with the well-established notion in the literature of PCR.
The relationship between these two approaches is discussed 
at length in \cite{Lin2020}, and also briefly in Section~\ref{sec:PCR} below.
In \cite{Lin2020}, we also analyzed the simpler cases of PGT$^+$
with vanishing torsion and curvature, respectively, which are not
merely special cases of the full PGT$^+$ Lagrangian, because
additional constraints are placed not only on Lagrangian
coefficients, but also on the fields. Although a number of unitary critical cases
were identified, no case was found that is also PPCR.

In seeking gravitational gauge theories that are renormalizable, one
promising route is to demand local scale invariance {\em a priori},
since such theories contain no dimensionful parameters, and hence no
absolute energy scale. Thus, rather than gauging the Poincar\'e group,
one may instead gauge the Weyl group, so that the action is also
invariant under local dilations. The resulting Weyl gauge theories
(WGT) were first discussed in
\cite{Bregman1973,Charap1974,Kasuya1975}.  In this article, we apply
our systematic method for identifying ghost-and-tachyon-free critical
cases to parity-preserving Weyl gauge theory (WGT$^+$), the
ground-state particle spectrum of which has rarely been discussed in
the literature before. 

This paper is arranged as follows. In Section II, we give a brief
introduction to WGT$^+$, and in Section III we consider the unitarity
of the `root' theory, where none of the critical conditions are
satisfied. In Section IV we apply our systematic approach to
investigating its critical cases and accommodating the associated
additional source constraints, as well as identifying some unitary
critical cases that are also propagating power-counting renormalizable.  We repeat
our analysis for WGT$^+$ with vanishing torsion in Section V and for
WGT$^+$ with vanishing curvature in Section VI. We conclude in Section VII.

We use the Landau--Lifshitz `mostly minus' metric signature $(+,-,-,-)$
throughout this paper.

\section{Weyl gauge theories}
\label{sec:wgt}

The action of an infinitesimal element of the Weyl group $W(1,3)$ on
Cartesian coordinates in Minkowski spacetime has the form
\begin{equation}
x^{\mu} \rightarrow x'^{\mu}=x^{\mu}+\epsilon^\mu + \omega\idx{^\mu_\nu}x^\nu+\rho x^\mu,
\end{equation}
where $\epsilon^\mu$ denotes a translation, $\omega\idx{^\mu_\nu}$
denotes a Lorentz rotation, and $\rho$ denotes a dilation. The
corresponding form variation
$\delta_0\varphi(x)\equiv\varphi'(x)-\varphi(x)$ of a field $\varphi$
(belonging to an irreducible representation of the Lorentz group) is
$\delta_0 \varphi = \delta_0^{\rm P}\varphi+ w\rho \varphi$, where
$\delta_0^{\rm P}$ means the variation under a Poincar\'e
transformation and $w$ is a dimensionless constant known as the (Weyl)
weight of the field.

One gauges the Weyl group $W(1,3)$ by demanding that the action be
invariant with respect to (infinitesimal, passively interpreted)
general coordinate transformations (GCTs) and the local action of the
subgroup $H(1,3)$ (the homogeneous Weyl group), obtained by setting
the translation parameters $\epsilon^\mu$ of $W(1,3)$ to zero (which
leaves the origin $x^\mu=0$ invariant), and allowing the remaining
group parameters to become independent arbitrary functions of
position.  In this way, one is led to the introduction of the
gravitational gauge fields ${h_A}^{\mu}$, ${A^{AB}}_{\mu}$ and
$B_\mu$, corresponding to the translational, rotational and dilational
parts of the Weyl group, respectively, which transform under the
gauged Weyl group as $\delta_0 h\idx{_A^\mu}=\delta_0^{\rm P}
h\idx{_A^\mu}-\rho h\idx{_A^\mu}$, $\delta_0
A\idx{^{AB}_\mu}=\delta_0^{\rm P} A\idx{^{AB}_\mu}$ and $\delta_0
B_\mu = -\partial_\mu\rho$.

The gauge fields are used to assemble the WGT covariant derivative 
\cite{Blagojevic2002,Lasenby2016}
\begin{equation}
\De^*_A\varphi = h\idx{_A^\mu} \De^*_\mu\varphi = h\idx{_A^\mu} \left(\de_{\mu} +
\tfrac{1}{2}A\idx{^{AB}_\mu}\Sigma_{AB} + wB_\mu\right)\varphi, 
\label{eqn:covderivdef}
\end{equation}
where $w$ is the weight of $\varphi$ and $\Sigma_{AB} = -\Sigma_{BA}$
are the generator matrices of the $\mbox{SL}(2,C)$ representation to
which $\varphi$ belongs.  The asterisk on the derivative operators is
a common notation used in WGT to distinguish these operators from
their PGT counterparts (to which they reduce if $w$ or $B_\mu$
vanishes). The corresponding commutators become
\begin{align*}
\left[\De^*_\mu,\De^*_\nu\right]\varphi=&\tfrac{1}{2}\mathcal{R}\idx{^{AB}_{\mu\nu}}\Sigma_{AB}\varphi+\mathcal{H}_{\mu\nu}w\varphi, \numberthis\\
\left[\De^*_A,\De^*_B\right]\varphi=&\tfrac{1}{2}\mathcal{R}\idx{^{CD}_{AB}}\Sigma_{CD}\varphi-\mathcal{T}\idx{^{*C}_{AB}}\De^*_C\varphi+\mathcal{H}_{AB}w\varphi,\numberthis
\end{align*}
where the field strengths have the forms
\begin{align}
\mathcal{R}\idx{^{AB}_{\mu\nu}} &= 
2(\de_{[\mu}A\idx{^{AB}_{\nu]}}+A\idx{^A_{E[\mu}}A\idx{^{EB}_{\nu]}}),\\
\mathcal{H}_{\mu\nu} &= 2\de_{[\mu}B_{\nu]}, \\
\mathcal{T}\idx{^{*C}_{AB}} &= \mathcal{T}\idx{^{C}_{AB}} + 2B_{[A} \delta^C_{B]},
\end{align}
and $\mathcal{T}\idx{^{C}_{\mu\nu}} = 2\De_{[\mu}{b^C}_{\nu]}$ is the
usual expression for the translational gauge field strength in PGT.
In the above expressions, Latin and Greek indices are related by
$h\idx{_A^\nu}$ and its inverse $b\idx{^A_\nu}$, with the relation
\begin{equation}
g_{\mu\nu} h\idx{_A^\mu} h\idx{_B^\mu} = \eta_{AB},\quad \eta_{AB} b\idx{^A_\mu} b\idx{^B_\mu} = g_{\mu\nu}.
\end{equation}
One may show that the weights of the translational and rotational
gauge fields are $w(h\idx{_A^\mu})=-1$ and $w(A\idx{^{AB}_\mu})=0$, so
that $w(b\idx{^A_\mu})=1$ and the weight of its determinant is
$w(b)=4$, but the dilatational gauge field $B_\mu$ itself
transforms inhomogeneously under dilations, as expected.  The weights
of the corresponding field strengths are
$w(\mathcal{R}\idx{^{CD}_{AB}})=w(\mathcal{H}_{AB})=-2$ and
$w(\mathcal{T}\idx{^{*C}_{AB}})=-1$.

In the action $S = \int
  b\mathcal{L}\,\di[4]{x}$, the Lagrangian $\mathcal{L}$ is the sum of
terms corresponding to the free gravitational fields and terms
containing the matter fields, respectively, and has the general form
\vspace*{-0.2cm}
\begin{equation}
\mathcal{L} = \mathcal{L}_{\rm
  G}(\mathcal{R}\idx{^{CD}_{AB}},\mathcal{T}\idx{^{*C}_{AB}},\mathcal{H}_{AB})+\mathcal{L}_{\rm
  M}(\varphi,\De^*_{A}\varphi). \label{eqn:WGTLag1}
\vspace{0.2cm}
\end{equation}
For $S$ to be scale invariant (i.e. of weight 0),
the weights of both $\mathcal{L}_{\rm G}$ and $\mathcal{L}_{\rm M}$ must
be $-4$. Restricting our attention to terms in $\mathcal{L}_F$ that
are at most quadratic in the field strengths, these may thus be
quadratic in $\mathcal{R}\idx{^{CD}_{AB}}$ and $\mathcal{H}_{AB}$, or
consist of the product of the two, but
may not include terms linear in $\mathcal{R}\idx{^{CD}_{AB}}$ or
quadratic in $\mathcal{T}\idx{^{*C}_{AB}}$.

One can, however, include further terms in the Lagrangian by
introducing an additional massless scalar field (or fields) $\phi$
with Weyl weight $w({\phi})=-1$, often termed the compensator(s)
\cite{Blagojevic2002}, which is usually non-minimally (conformally)
coupled to the field strength tensors of the gravitational gauge
fields. For example, terms proportional to $\phi^2 {\cal R}$ or
$\phi^2 \mathcal{L}_{{\cal T}^{\ast 2}}$, where $\mathcal{L}_{{\cal
    T}^{\ast 2}}$ consists of terms quadratic in ${{\cal T}^{\ast
    C}}_{AB}$, have weight $w=-4$ and so may be added to the total
Lagrangian \cite{Dirac1973,Omote1977,Sijacki1982,Neeman1988}. One should also
include a free kinetic term $(\De^\ast\phi)^2$ for the scalar field,
and may also add a self-interaction term $\phi^4$, but we shall not
consider the latter here. Thus, also requiring parity-invariance, 
the Lagrangian for free WGT$^+$ has the form
\begin{widetext}
\begin{align*}
\mathcal{L}_\text{G}=&
-\lambda \phi^2 \mathcal{R}
+\tfrac{1}{6}\left(2 r_1 + r_2\right) 
\mathcal{R}^{ABCD}\mathcal{R}_{ABCD}
+\tfrac{2}{3}\left(r_1-r_2\right)
\mathcal{R}^{ABCD} \mathcal{R}_{ACBD}
+\tfrac{1}{6} \left(2 r_1+r_2-6 r_3\right)
\mathcal{R}^{ABCD} \mathcal{R}_{CDAB} \\[0.1cm]
&+\left(r_4+r_5\right)
\mathcal{R}^{AB} \mathcal{R}_{AB}
+\left(r_4-r_5\right) 
\mathcal{R}^{AB} \mathcal{R}_{BA} 
+c_1 \mathcal{R}^{AB} \mathcal{H}_{AB}
+\xi \mathcal{H}^{AB} \mathcal{H}_{AB}
+\tfrac{1}{2}\nu \De^*_A\phi \De^{*A}\phi\\[0.1cm]
&+\tfrac{1}{12} \left(4 t_1+t_2+3 \lambda \right) \phi^2
\mathcal{T}^{*ABC} \mathcal{T}^*_{ABC}
-\tfrac{1}{6} \left(2 t_1-t_2+3 \lambda \right) \phi^2
\mathcal{T}^{*ABC} \mathcal{T}^*_{BCA}
-\tfrac{1}{3} \left(t_1-2 t_3+3 \lambda \right) \phi^2
\mathcal{T}\text{}^*_B\text{}^A\text{}^B \mathcal{T}\text{}^*_{CA}\text{}^C,
\label{eqn:WGTLag}\numberthis
\end{align*}
\end{widetext}
where $\mathcal{R}\idx{^A_B}=\mathcal{R}\idx{^{AC}_{BC}}$,
$\mathcal{R}=\mathcal{R}\idx{^A_A}$ and $\De^*_A\phi = \de_A \phi -
B_A\phi$. The parameters in the Lagrangian are dimensionless and set
in combinations that enable a straightforward comparison with our
previous studies of PGT$^+$ \cite{Lin2019a,Lin2020}. Note that the
Gauss--Bonnet identity has been used to remove the term proportional
to $\mathcal{R}^2$.

Provided $\phi(x)$ does not vanish anywhere, one can use local scale
invariance to set the field to a constant value $\phi_0$, which is
known as the Einstein gauge and is usually interpreted as breaking the
scale symmetry. This interpretation is questioned in
\cite{Lasenby2016}, however, since it is shown that if one rewrites
the Lagrangian in terms of a set of scale-invariant variables
\cite{Kasuya1975}, then the resulting equations of motion are the same
as those of Einstein gauge, yet this approach involves no breaking of
the scale symmetry.
In any case, we will adopt the Einstein gauge $\phi=\phi_0$ here, the
most significant effect of which is that the term $\frac{1}{2}\nu
\De^*_A\phi\De^{*A}\phi$ in the Lagrangian becomes $\frac{1}{2}\nu
\phi_0^2 B_A B^A$. We then absorb the $\phi_0^2$ factor into the
now dimensionful parameters $\lambda$, $\nu$, $t_1$, $t_2$, and
$t_3$, without loss of generality.
Note that a potential term $\sim\phi^4$ for the compensator
scalar field was not included in the Lagrangian, since it becomes a
constant in the Einstein gauge, acting like an effective cosmological
constant, which would be inconsistent with considering a Minkowski
background.

WGT is most naturally interpreted as a field theory in Minkowski
spacetime\cite{Wiesendanger1996,Lasenby1998,Lasenby2016}, in the same way as
the gauge field theories describing the other fundamental
interactions.  It is more common, however, to reinterpret it
geometrically in terms of a Weyl--Cartan spacetime ($W_4$), which
generalises the Riemann--Cartan spacetime ($U_4$) underlying the
geometric interpretation of PGT by incorporating local scale
invariance \cite{Blagojevic2002}.


Weyl--Cartan spacetime is a manifold with linear connection ($\Gamma$)
and metric ($g_{\mu\nu}$) which satisfy
\begin{equation}
\De^*_\rho(\Gamma) g_{\mu\nu} = 0, \label{eqn:WGTSemiMetricity}
\end{equation}
where the covariant derivative of a field $\varphi$ with weight $w$ is
defined by
\begin{equation}
\De^*_\mu(\Gamma) \varphi \equiv \left(\De_\mu(\Gamma)+wB_\mu\right)
\varphi, 
\label{eqn:WGTCovDev}
\end{equation}
in which $\De_\mu(\Gamma) = \partial_\mu + {\Gamma^\sigma}_{\rho\mu}
{\matri{X}^\rho}_\sigma$ is the $U_4$ covariant derivative and
${\matri{X}^\rho}_\sigma$ are the $\mbox{GL}(4,R)$ generator matrices
appropriate to the GCT tensor character of the field to which the
operator is applied.
The semi-metricity condition \eqref{eqn:WGTSemiMetricity} replaces the
metricity condition in $U_4$. Since $w(g_{\mu\nu})=2$, the
semi-metricity condition can also be written as $\De_\rho(\Gamma)
g_{\mu\nu} = -2B_\rho g_{\mu\nu}$, from which one finds that the
infinitesimal change of length of a parallel transported vector is
proportional to the length itself, $\De_\rho(\Gamma) V^2 =
-2B_\rho V^2$. One may solve for the connection $\Gamma$, which is
given by
\begin{equation}
\Gamma\idx{^\mu_{\nu\rho}} =\Christoffel{\mu}{\nu\rho}+\delta^\mu_\nu B_\rho +\delta^\mu_\rho B_\nu - g_{\nu\rho}B^\mu +K\idx{^\mu_{\nu\rho}}, 
\end{equation}
where $\Christoffel{\mu}{\nu\rho}$ is the ordinary Christoffel symbol
and $K\idx{^\mu_{\nu\rho}}$ is the contorsion tensor (discussed
further below).

A local Lorentz frame at each point on the manifold describes the
tangent space and is determined by the tetrad basis $h\idx{_A^\mu}$
with its inverse $b\idx{^A_\mu}$; these
quantities may be used to convert between coordinate and local Lorentz
indices.
The Minkowski metric $\eta_{AB}$ is invariant under Weyl
transformation, so $w(\eta_{AB})=0$ and $w(h\idx{_A^\mu})=-1$. The
local frame has a connection $A\idx{^{AB}_\mu}$, and the covariant
derivative $\De^*_A(A)$ has properties similar to
\eqref{eqn:WGTCovDev}, where
\begin{align}
&\De^*_\rho(A) \eta_{AB} = 0, \\ &\De^*_\rho(A) \varphi \equiv
  \left(\De_\rho(A)+w B_\rho\right) \varphi,
\end{align}
and $\De_\rho(A)$ is the covariant derivative in $U_4$. One may also
define the `total covariant derivative' $\De^*_\rho(\Gamma+A)$ to act 
on quantities
with both coordinate and local Lorentz indices
\begin{equation}
\De^*_\rho(\Gamma+A)\varphi =  \left(\De_\rho(\Gamma)+\De_\rho(A)-\de_\rho-w B_\rho\right) \varphi.
\end{equation}
Since the total covariant derivative $\De^*_\rho(\Gamma+A)V^A$ of the
local Lorentz components of a vector is a coordinate tensor in
Weyl--Cartan spacetime, the relation $\De^*_\rho (\Gamma+A)V^A =
b\idx{^A_\mu}\De^*_\rho(\Gamma+A) V^\mu$ should hold, from which one
obtains the so-called `tetrad postulate'
\begin{equation}
D^*_\mu(\Gamma+A)b\idx{^A_\nu}\equiv
\de_{\mu}^*b\idx{^A_\nu}+A\idx{^A_{B\mu}}b\idx{^B_\nu}-\Gamma\idx{^\sigma_{\nu\mu}}b\idx{^A_\sigma}=0,
\end{equation}
where $\partial^\ast_\mu \equiv \partial_\mu + w B_\mu$.
One can therefore express the affine connection in the quantities
corresponding to gauge fields as
\begin{equation}
\Gamma\idx{^\lambda_{\nu\mu}}=h\idx{_A^\lambda}(\de_{\mu}^*b\idx{^A_\nu}+A\idx{^A_{B\mu}}b\idx{^B_\nu}), \label{eqn:WGTConnectionGaugeFields}
\end{equation}
and hence show that the translational gauge field strength
is equivalent to (minus) the geometric torsion tensor
\begin{equation}
\mathcal{T}\idx{^{*\rho}_{\mu\nu}} 
= \Gamma\idx{^\rho_{\nu\mu}}-\Gamma\idx{^\rho_{\mu\nu}}, 
\label{eqn:WGTTorsionConnection}
\end{equation}
in terms of which the contorsion is given by
\begin{equation}
K_{\mu\lambda\nu}
=-\tfrac{1}{2}\left(\mathcal{T}_{\mu\lambda\nu}-\mathcal{T}_{\nu\mu\lambda}+\mathcal{T}_{\lambda\nu\mu}\right). \label{eqn:WGTContorsionTorsion}
\end{equation}

From \eqref{eqn:WGTConnectionGaugeFields},
\eqref{eqn:WGTTorsionConnection}, and
\eqref{eqn:WGTContorsionTorsion}, one also obtains
\begin{equation}
A_{AB\mu}=\Delta^*_{AB\mu}+K_{AB\mu}, \label{eqn:WGTADeltaK}
\end{equation}
where we define the quantities
\begin{align}
\Delta^*_{AB\mu}&\equiv\Delta_{AB\mu}|_{\de\to\de^*} = \Delta_{AB\mu}-B_A b_{B\mu}+B_B b_{A\mu},\\
\Delta_{AB\mu}&\equiv \frac{1}{2}\left(c_{ABC}-c_{CAB}+c_{BCA}\right)b\idx{^C_\mu},\\
c\idx{^A_{\mu\nu}}&\equiv \de_{\mu}b\idx{^A_\nu}-\de_{\nu}b\idx{^A_\mu}.
\end{align}
One then finds that, in contrast to the torsion, the geometric
(Riemann) curvature tensor differs from the rotational gauge field
strength $\mathcal{R}\idx{^\rho_{\sigma\mu\nu}}$, so we denote the
former by
\begin{align*}
\tilde{\mathcal{R}}\idx{^\rho_{\sigma\mu\nu}}&=\mathcal{R}\idx{^\rho_{\sigma\mu\nu}} + H_{\mu\nu}\delta^\rho_\sigma, \\
&=\de_{\mu}\Gamma\idx{^\rho_{\sigma\nu}}-\de_{\nu}\Gamma\idx{^\rho_{\sigma\mu}}+\Gamma\idx{^\rho_{\lambda\mu}}\Gamma\idx{^\lambda_{\sigma\nu}}-\Gamma\idx{^\rho_{\lambda\nu}}\Gamma\idx{^\lambda_{\sigma\mu}}. \numberthis
\end{align*}
Unlike $\mathcal{R}\idx{_{\rho\sigma\mu\nu}}$, the curvature tensor
$\tilde{\mathcal{R}}\idx{_{\rho\sigma\mu\nu}}$ is not antisymmetric in
$(\rho,\sigma)$, while both are antisymmetric in $(\mu,\nu)$
\cite{Blagojevic2002,Lasenby2016}. Indeed, one may take advantage of
these symmetry properties by using
$\mathcal{R}\idx{_{\rho\sigma\mu\nu}}$ to perform calculations instead
of $\tilde{\mathcal{R}}\idx{_{\rho\sigma\mu\nu}}$. One should note,
however, that unlike the curvature tensor in Riemann spacetime $V_4$
familiar from general relativity, neither $\mathcal{R}\idx{_{\rho\sigma\mu\nu}}$ nor
$\tilde{\mathcal{R}}\idx{_{\rho\sigma\mu\nu}}$ is symmetric in
$(\rho\sigma,\mu\nu)$.

\section{The `root' theory}
\label{sec:root}

We now apply the method described in \cite{Lin2019a} to the `root'
theory \eqref{eqn:WGTLag}, where none of the critical conditions is
satisfied. We first linearize the Lagrangian around the Minkowski
background using $A_{ABC}\sim O(t)$, $B_{A}\sim O(t)$,
$h\text{}_A\text{}^{\mu }=\delta\text{}_A\text{}^{\mu
}+f\text{}_A\text{}^{\mu }$, and
$f\text{}_A\text{}_B=\mathfrak{s}\text{}_A\text{}_B-
\mathfrak{a}\text{}_A\text{}_B\sim O(t)$, where $\mathfrak{s}$ and
$\mathfrak{a}$ denote the symmetric and antisymmetric parts of $f$,
respectively.
Note that we
cannot perturb $\phi$ as $\phi_0+\epsilon$, for some excitation
$\epsilon$, because we have already fixed the gauge on $\phi$.  The
Lagrangian then becomes
\begin{equation}
b\mathcal{L}_{\rm G} = -\left(2\lambda \de_A A\idx{^B^A_B}\right) + \mathcal{O}\left(t^2\right),
\label{eqn:WGTFirstOrder}
\end{equation}
where the linear term is just a total derivative. We then decompose
the quadratic part into
	\begin{align}b\mathcal{L}_\mathrm{G} = \sum_{J,P,i,j}a(J^P)_{ij}\hat{\zeta}^\dagger\cdot \hat{P}(J^P)_{ij}\cdot \hat{\zeta},
\label{eqn:Ldecomp}
	\end{align}
	using the spin projection operators (SPOs) $\hat{P}(J^P)_{ij}$
        \cite{Rivers1964,Barnes1965,Aurilia1969}. Section II of
        \cite{Lin2019a} contains a description of our notation (note
        that Eq. (52) in \cite{Lin2019a} contains a typographical
        error and should read
        $f\text{}_A\text{}_B=\mathfrak{s}\text{}_A\text{}_B-
        \mathfrak{a}\text{}_A\text{}_B$, as here, but this does not
        affect the remaining contents in \cite{Lin2019a,Lin2020}). The
        SPOs for WGT$^+$ are given in
        \Cref{sec:SpinProjectionOperatorWGT}. One then obtains the
        $a$-matrices:
\begin{widetext}
\begin{align}
a(0^-)=&\bordermatrix{
	~&A \cr
A&2 \left(k^2 r_2+t_2\right) \cr
}, \label{eq:azerominus}\\[0.2cm]
a(0^+)=&\bordermatrix{
	~&A &\mathfrak{s} &\mathfrak{s} &B \cr
A& 2 \left(2 k^2 \left(r_1-r_3+2 r_4\right)+t_3\right) & -2 i \sqrt{2} k t_3 & 0 & 2 \sqrt{6} \left(t_3-\lambda \right) \cr
\mathfrak{s}& 2 i \sqrt{2} k t_3 & 4 k^2 \left(t_3-\lambda \right) & 0 & 4 i \sqrt{3} k \left(t_3-\lambda \right) \cr
\mathfrak{s}&0 & 0 & 0 & 0 \cr
B&2 \sqrt{6} \left(t_3-\lambda \right) & -4 i \sqrt{3} k \left(t_3-\lambda \right) & 0 & 4 \left(3 t_3-3 \lambda +\frac{\nu }{4}\right) \cr
}, \\[0.2cm]
a(1^-)=&\bordermatrix{
	~&A &A &\mathfrak{s} &\mathfrak{a} &B \cr
A& \makecell[cl]{2 \left[k^2 \left(r_1+r_4+r_5\right)\right.\\+\left.\frac{1}{6} \left(t_1+4 t_3\right)\right]} & -\frac{\sqrt{2}}{3} \left(t_1-2 t_3\right) & \frac{\sqrt{2}}{3} i  k \left(t_1-2 t_3\right) & -\frac{\sqrt{2}}{3} i  k \left(t_1-2 t_3\right) & c_1 k^2-4 t_3+4 \lambda \cr
A& -\frac{\sqrt{2}}{3}  \left(t_1-2 t_3\right) & \frac{2}{3} \left(t_1+t_3\right) & -\frac{2}{3} i k \left(t_1+t_3\right) & \frac{2}{3} i k \left(t_1+t_3\right) & 2 \sqrt{2} \left(-t_3+\lambda \right) \cr
\mathfrak{s}&-\frac{\sqrt{2}}{3} i  k \left(t_1-2 t_3\right) & \frac{2}{3} i k \left(t_1+t_3\right) & \frac{2}{3} k^2 \left(t_1+t_3\right) & -\frac{2}{3} k^2 \left(t_1+t_3\right) & -2 i \sqrt{2} k \left(t_3-\lambda \right) \cr
\mathfrak{a}&\frac{\sqrt{2}}{3} i  k \left(t_1-2 t_3\right) & -\frac{2}{3} i k \left(t_1+t_3\right) & -\frac{2}{3} k^2 \left(t_1+t_3\right) & \frac{2}{3} k^2 \left(t_1+t_3\right) & 2 i \sqrt{2} k \left(t_3-\lambda \right) \cr
B& c_1 k^2-4 t_3+4 \lambda & 2 \sqrt{2} \left(-t_3+\lambda \right) & 2 i \sqrt{2} k \left(t_3-\lambda \right) & -2 i \sqrt{2} k \left(t_3-\lambda \right) & 4 \left(3 t_3-3 \lambda +\frac{\nu }{4}+ k^2 \xi \right) \cr
}, \\[0.2cm]
a(1^+)=&\bordermatrix{
	~&A &A &\mathfrak{a} \cr
A& \frac{1}{3} \left(6 k^2 \left(2 r_3+r_5\right)+t_1+4 t_2\right) & \frac{1}{3} \sqrt{2} \left(t_1-2 t_2\right) & \frac{1}{3} i \sqrt{2} k \left(t_1-2 t_2\right) \cr
A& \frac{1}{3} \sqrt{2} \left(t_1-2 t_2\right) & \frac{2}{3} \left(t_1+t_2\right) & \frac{2}{3} i k \left(t_1+t_2\right) \cr
\mathfrak{a}& -\frac{1}{3} i \sqrt{2} k \left(t_1-2 t_2\right) & -\frac{2}{3} i k \left(t_1+t_2\right) & \frac{2}{3} k^2 \left(t_1+t_2\right) \cr
}, \\[0.2cm]
a(2^-)=&\bordermatrix{
	~&A \cr
A& 2 \left(k^2 r_1+\frac{t_1}{2}\right) \cr
}, \\[0.2cm]
a(2^+)=&\bordermatrix{
	~&A &\mathfrak{s} \cr
A& 2 \left(k^2 \left(2 r_1-2 r_3+r_4\right)+\frac{t_1}{2}\right) & -i \sqrt{2} k t_1 \cr
\mathfrak{s}& i \sqrt{2} k t_1 & 2 k^2 \left(t_1+\lambda
\right). \label{eq:atwoplus} \cr
}
\end{align}
\end{widetext}

In general, if any of the matrices $a(J^P)$ in the decomposition
(\ref{eqn:Ldecomp}) are singular, then the theory possesses gauge
invariances. One may fix these gauges by deleting rows and columns of
the $a$-matrices such that they become non-singular. The elements of
the resulting matrices are usually denoted by $b_{ij}(J^P)$. 
For WGT$^+$, some of the $a$-matrices given above are indeed
singular. In particular, one may delete the third row/column of
$a(0^+)$, the fourth row/column of $a(1^-)$, and the third row/column
of $a(1^+)$ to obtain the corresponding non-singular $b$-matrices.
The singular nature of these three $a$-matrices results in them having
both null right and left eigenvectors, which give us gauge invariance and source constraints respectively.
For each spin-parity sector, the null left eigenvectors are given
by
\begin{align}
0^+:& \left(0,0,1,0\right) \\
1^-:& \left(0,ik,0,1,0\right),\left(0,-ik,1,0,0\right) \\
1^+:& \left(0,ik,1\right),
\end{align}
where one should note that the $B$-field is not involved, since the
corresponding vector component is always zero, and the remaining
components are the same as those found for PGT$^+$. This is no
surprise, since the dilation gauge invariance has been fixed by
adopting the Einstein gauge, and the remaining symmetry should indeed
be local Poincar\'e invariance.

The null eigenvectors may be used to derive the form of the associated
gauge invariances and the corresponding source constraints for
WGT$^+$, which are found to be the same as those in PGT$^+$, as
expected.  The gauge invariances are given by
\begin{align}
\delta h_{AB} &= u_{[AB]} + k_B v_A\\
\delta A_{ABC} &= -i k_C u_{[AB]},
\end{align}
where $u_{[AB]}$ and $v_A$ are some arbitrary fields, and 
the source constraints have the form
\begin{align}
k^A\sigma_{AB}&=0 \\
ik^A\tau_{ABC}+\sigma_{[BC]}&=0,
\end{align}
where $\sigma_{AB}$ is the source current of $f_{AB}$, and
$\tau_{ABC}$ is the source current of $A_{ABC}$. 

The requirement that a theory is free from ghosts and tachyons places
conditions on the $b$-matrices, and one must consider the massless and
massive particle sectors separately. For the massless modes, one
requires only that there be no ghosts. As discussed in
\cite{Lin2019a}, this is determined by considering the coefficient
matrices $\mathbf{Q}_{2n}$ in a Laurent series expansion of the
saturated propagator about the origin in momentum space. For WGT$^+$,
one finds that all of the entries $\mathbf{Q}_{2n}$ vanish identically
for $n>1$, and so the saturated propagator does not have a higher pole
at $k^2=0$. The non-zero eigenvalues of $\mathbf{Q}_{2}$ are found to
be
\begin{equation}
\frac{1+6 |\vec{k}|^2}{\lambda },\frac{1+8 |\vec{k}|^2}{2 \lambda }, \numberthis
\end{equation}
and so there are 2 degrees of freedom in the propagating massless
particle sector.\footnote{Note that the expression for the
  eigenvalues is not unique, but depends on the form
  chosen for the source constraints. To be precise, one can obtain another
  set of the null vectors $\mathbf{n}_i$ in Eq. (30) of
  \cite{Lin2019a} by linear combination.}  The massless no-ghost
condition is that all eigenvalues of $\mathbf{Q}_{2n}$ are
non-negative, and so one requires simply that
\begin{equation}
\lambda >0.
\end{equation}

Turning to the massive particle sector, one must first determine the
particle masses by calculating the determinants of the $b$-matrices:
\begin{align*}
\mathrm{det}\left[b\left(0^-\right)\right]=&2 k^2 r_2+2 t_2,
\numberthis\\
\mathrm{det}\left[b\left(0^+\right)\right]=&16 \left(r_1-r_3+2 r_4\right) \left(t_3-\lambda \right) \nu  k^4 \\&- 8\lambda\left[12 \left(t_3-\lambda \right) \lambda + t_3\nu\right]  k^2,
\numberthis \label{eqn:WGTdet0p}\\
\mathrm{det}\left[b\left(1^-\right)\right]=&-\tfrac{2}{3} \left(t_1+t_3\right) \left[c_1^2-8 \left(r_1+r_4+r_5\right) \xi \right]  k^4 \\
&+  \tfrac{4}{3} \left\{6 c_1 t_1 \left(t_3-\lambda \right)+\left(r_1+r_4+r_5\right)\right.\\ &\left.\left[12 \left(t_3-\lambda \right) \left(t_1+\lambda \right)+\left(t_1+t_3\right) \nu \right]\right.\\&\left.+6 t_1 t_3 \xi \right\}  k^2 +2 t_1 \left[12 \lambda \left(t_3-\lambda\right)+t_3 \nu \right],
\numberthis \label{eqn:WGTdet1n}\\
\mathrm{det}\left[b\left(1^+\right)\right]=&\frac{4}{3} \left(2 r_3+r_5\right) \left(t_1+t_2\right)  k^2  +  2 t_1 t_2,
\numberthis\\
\mathrm{det}\left[b\left(2^-\right)\right]=&2 r_1  k^2  + t_1,
\numberthis\\
\mathrm{det}\left[b\left(2^+\right)\right]=&4 \left(2 r_1-2 r_3+r_4\right) \left(t_1+\lambda \right)  k^4  +  2 t_1 \lambda  k^2, \numberthis
\end{align*}
from which one finds that there is no massive mode in the $0^+$
sector, and the particle masses in the other sectors are given by
\begin{align}
m^2\left(0^-\right)&=-\frac{t_2}{r_2}, \label{eqn:msq0+}\\
m^2\left(0^+\right)&=\frac{12 \lambda^2\left(t_3-\lambda \right) + t_3\lambda}{2 \left(r_1-r_3+2 r_4\right) \left(t_3-\lambda \right) \nu}, \\
m^2\left(1^-\right)&=\text{(the two roots of $\mathrm{det}\left[b\left(1^-\right)\right]$)},\\
m^2\left(1^+\right)&=-\frac{3 t_1 t_2}{2 \left(2 r_3+r_5\right) \left(t_1+t_2\right)},\\
m^2\left(2^-\right)&=-\frac{t_1}{2 r_1},\\
m^2\left(2^+\right)&=-\frac{t_1 \lambda }{2 \left(2 r_1-2 r_3+r_4\right) \left(t_1+\lambda \right)}.\label{eqn:msq2+}
\end{align}
The no-tachyon conditions are then simply $m^2(J^P)>0$. We give the
conditions for the $1^-$ sector in \Cref{sec:MassiveCond1nWGT} because
of the length of the expressions involved. Note also for the $1^-$
sector that one requires the two roots of \eqref{eqn:WGTdet1n} to be
distinct in order to avoid a dipole ghost. Hence, in each sector, the
masses are distinct, and so one can apply Eq. (45) in \cite{Lin2019a}
directly to obtain the massive no-ghost conditions:
\begin{align*}
0^-:& r_2<0, \numberthis
\\
0^+:& \left(r_1-r_3+2 r_4\right) \left(t_3-\lambda \right) \lambda  \nu ^2 \left\{24 \left(t_3-\lambda \right) \lambda ^3 \right.\\&\left.+12 \left(r_1-r_3+2 r_4\right) \left(t_3-\lambda \right) \lambda  \nu +\left[\left(r_1-r_3+2 r_4\right) t_3\right.\right.\\&\left.\left.+t_3 \lambda -\lambda ^2\right] \nu ^2\right\}>0, \numberthis
\\
1^+:&\left(2 r_3+r_5\right) >0, \numberthis
\\
2^-:& r_1<0, \numberthis
\\
2^+:& \lambda  \left(2 r_1-2 r_3+r_4\right) \left(\lambda +t_1\right) \\
&\hphantom{r_1+}\left[\left(2 r_1-2 r_3+r_4\right) t_1-\lambda ^2-\lambda  t_1\right]<0, \numberthis
\end{align*}
where again we do not write out the condition for $1^-$ because of its
length, but instead give the relevant expression in
\Cref{sec:MassiveCond1nWGT}.

The combined no-ghost-and-tachyon conditions for each sector other
than $1^-$ are then
\begin{align}
&0^-:t_2>0,\; r_2<0
\\
&0^+:r_1+2 r_4>r_3,\; \left(t_3-\lambda \right) \lambda  \nu  \left[12 \lambda\left(t_3-\lambda\right)+t_3 \nu\right]>0
\\
&1^+:2 r_3+r_5>0,\; t_1 t_2 (t_1+t_2)<0
\\
&2^-:t_1>0, r_1<0
\\
&2^+:2 r_1+r_4>2 r_3,\; \lambda t_1 (\lambda+t_1)<0.
\end{align}
For the $1^-$ sector, we give the combined condition in
\Cref{sec:MassiveCond1nWGT} and show that it does allow some ranges of
the parameters, but we are unable to obtain a simplified expression
for it. Note that, except for the $0^+$ and $1^-$ sectors, the
combined condition in each of the other spin-parity sectors is exactly
the same as originally found in \cite{Sezgin1980} for PGT$^+$.

Finally, if we consider all the no-tachyon and no-ghost conditions from
all the massive sectors, we find that they cannot be satisfied
simultaneously. Thus, the root theory must contain a massive ghost or
tachyon.

\section{Critical cases}

If the parameters in the action satisfy certain `critical conditions',
the particle masses (\ref{eqn:msq0+})--(\ref{eqn:msq2+}) can become
zero or infinite, and the resulting critical cases may
possess additional gauge invariances, so one may have to re-evaluate the
no-tachyon and no-ghost conditions for both the massless and massive
sectors.

\subsection{Unitarity}

In attempting to apply the method in \cite{Lin2019a} to obtain all the
critical cases of the root WGT$^+$ theory, one finds that some of the
coefficients in \Cref{eqn:WGTdet0p,eqn:WGTdet1n} cannot be factorized
into linear combinations of the parameters. Consequently, the method
in \cite{Lin2019a} cannot be applied straightforwardly to obtain all
the critical cases, and one must check carefully where it is
applicable.  For example, one of the factors in the coefficient of the
$k^2$ term in \eqref{eqn:WGTdet0p} is
\begin{equation}
12 \left(t_3-\lambda \right) \lambda + t_3\nu, \label{eqn:WGTNonLinearCoeff}
\end{equation}
which cannot be written as the product of factors that are linear in
the Lagrangian parameters. Indeed, for (\ref{eqn:WGTNonLinearCoeff})
to equal zero, one has the two solutions:
\label{eqn:WGTNonLinearSol}
\begin{align}
&\nu=-\frac{12 \left(t_3-\lambda \right) \lambda}{t_3}\quad \text{ with }\; t_3\neq 0, \label{eqn:WGTNonLinearSolA}\\
&t_3=\lambda=0.\label{eqn:WGTNonLinearSolB}
\end{align}
It is therefore not as straightforward to apply the condition $12
\left(t_3-\lambda \right) \lambda + t_3\nu=0$ by
substitution. Moreover, the second solution
\eqref{eqn:WGTNonLinearSolB} requires one to eliminate two degrees of
freedom in the parameters simultaneously and thus breaks the hierarchy
of the `tree' of critical cases discussed in \cite{Lin2019a}. 

In general, one finds that allowing any of the Lagrangian parameters
$\nu$, $\xi$, or $c_1$ in \eqref{eqn:WGTLag} to be non-zero
introduces similar problems.  It requires further improvement of our
systematic method to accommodate such cases, and so here we set
$\nu=\xi=c_1=0$ to avoid these difficulties. Thus, for the remainder
of this section, the `root theory' refers to \eqref{eqn:WGTLag} with
$\nu=\xi=c_1=0$. As we will show below, however, one may nevertheless
construct a theory with $\nu\neq 0$ and/or $\xi\neq 0$ from a theory with
$\nu=\xi=0$, provided its $a$-matrices are `non-mixing'.

Starting from the `root' theory, we systematically identify
\numCriticalNotRootNotZero{} critical cases (excluding the `vanishing'
Lagrangian, for which all parameters are zero). Of these critical
cases, we find \numNoGhostNoTachyon{} are free of ghosts and tachyons,
provided the parameters in each case satisfy some additional
conditions that preclude them from generating another critical case;
this general issue is discussed in detail in \Cref{sec:completeness}. 
The full set of results, displayed in an interactive form, can be found
at: \url{http://www.mrao.cam.ac.uk/projects/gtg/wgt/}.

\subsection{Comparison with previous results}

We now compare our results with the only other
example of a unitary WGT$^+$ theory of which we are aware in the
literature \cite{Nieh1982}. This has the Lagrangian
\begin{equation}
\mathcal{L} = -\lambda\phi^2\mathcal{R} +a\mathcal{R}^2-\tfrac{1}{4}H^{\mu\nu}H_{\mu\nu}+\tfrac{1}{2}\De^*_\mu\phi \De^{*\mu}\phi,
\label{eqn:Niehtheory}
\end{equation} 
which on adopting the Einstein gauge becomes
\begin{equation}
\mathcal{L} = -\lambda\phi_0^2\mathcal{R}
+a\mathcal{R}^2-\tfrac{1}{4}H^{\mu\nu}H_{\mu\nu}+\tfrac{1}{2}\phi_0^2
B_\mu B^\mu.
\end{equation} 
Thus, the $B$-field is decoupled from the other gauge fields and so
the theory can be viewed as the combination of PGT$^+$ with
$\mathcal{L} = -\lambda\phi_0^2\mathcal{R} +a\mathcal{R}^2$ and
Proca theory $\mathcal{L}_{\rm Pr} =
-\frac{1}{4}H^{\mu\nu}H_{\mu\nu}+\frac{1}{2}\phi_0^2 B_\mu B^\mu$ for
a massive vector field. The Proca part is well-known to be unitary.
Using the Gauss--Bonnet identity, the PGT$^+$ part may be shown to
correspond to the critical case
$r_1=r_2=2r_3-r_4=2r_3+r_5=t_1+t_2=t_1+t_3=t_1+\lambda=0, r_3\neq0,
\lambda\neq0$. This a type C critical case of the root PGT$^+$ theory
with no massive mode and massless modes with 2 degrees of freedom; the
no-ghost-and-tachyon condition is simply $\lambda>0$. Therefore,
provided this condition is satisfied, the theory
(\ref{eqn:Niehtheory}) is indeed unitary.

One should note that the presence of the kinetic terms for the $B$ and
$\phi$ fields means that (\ref{eqn:Niehtheory}) is not a critical case
of our redefined WGT$^+$ with $\nu=\xi=c_1=0$ in
(\ref{eqn:WGTLag}), but is a critical
case of the `full' WGT$^+$ root theory without this
constraint on the Lagrangian parameters. In particular,
(\ref{eqn:Niehtheory}) belongs to an extended set of theories with
$\nu\neq 0$ and $\xi\neq 0$ that can be separated into a PGT$^+$ part
and a dilaton part, which we discuss
below in
the context of propagating power-counting renormalizability. We
  note, however, that the PGT$^+$
part of (\ref{eqn:Niehtheory}) is not listed in \cite{Lin2020} because
one cannot obtain non-mixing $b$ matrices 
by deleting
rows and columns from its $a$ matrices.
 
\subsection{Propagating power-counting renormalizability}
\label{sec:PCR}

In addition to possessing no ghosts or tachyons, a healthy physical
theory should also be renormalizable. The first step in assessing
whether this is possible is to determine whether the theory is
power-counting renormalizable (PCR).

As discussed in \cite{Lin2019a,Lin2020}, the key quantity for
determining whether a theory is PCR is the propagator
        	\begin{equation}
        	\hat{D} = \sum_{J,P,i,j} b^{-1}_{ij}\hat{P}(J^P)_{ij}.
        	\end{equation}
In particular, if the $b$-matrices are block diagonal, with each block
containing only one of the fields $A$, $\mathfrak{s}$, $\mathfrak{a}$
and $B$, then there are no mixing terms in the (gauge-fixed)
Lagrangian and it is straightforward to obtain the propagators for
these fields separately from $\hat{D}$. Extending the original PCR
criterion used by Sezgin in \cite{Sezgin1980} would require the
propagator of the $A$ and $B$ fields to decay at least as quickly as
$k^{-2}$, respectively, at high energy, and those of the
$\mathfrak{s}$ and $\mathfrak{a}$ fields to fall off at least as
$k^{-4}$ (see \Cref{sec:PCRdef}). By contrast, we proposed an
alternative criterion in \cite{Lin2019a,Lin2020}, which we now term
propagating power counting renormalizability (PPCR), that in addition
allows the presence of non-propagating fields at high momenta (for
which the propagator decays no faster than $\sim k^0$).  Since the
physical basis of power-counting renormalizability relates to the
divergence at large momenta of integrals describing the propagation of
particles around closed loops in Feynman diagrams, it seems physically
reasonable to allow for the presence of modes that do not propagate at
large momenta, since these should be integrated out and not contribute
to the loop integrals. PPCR is less restrictive than PCR, and it may
therefore retain some theories that are eliminated by PCR erroneously.
The ultimate consistency of these two approaches in identifying
particular theories as PCR and PPCR is discussed at length in
\cite{Lin2020}, although the second approach is preferred since it
identifies further critical cases that reduce to those identified by
Sezgin's criterion at linear level after integrating out any
non-propagating modes.  We therefore again adopt the latter method
here, which is consistent with our previous work.

On performing this analysis, one finds that most of the critical cases
identified as PPCR are identical to those listed in Table
I, III or V in \cite{Lin2020}, or are a PGT$^+$ without any
propagating mode (which were not listed in \cite{Lin2020}) but with an
additional propagating dilaton.  One may understand the reason for
this by first expanding the $\mathcal{T}^{*2}$ terms in
\eqref{eqn:WGTLag} to obtain
\begin{align}
\mathcal{T}^{*}_{ABC}\mathcal{T}^{*ABC} &= \mathcal{T}_{ABC}\mathcal{T}^{ABC} + 4 B_A T^{CA}_C + 6 B^A B_A, \\
\mathcal{T}^{*}_{ABC}\mathcal{T}^{*BCA} &= \mathcal{T}_{ABC}\mathcal{T}^{BCA} -2 B_A T^{CA}_C -3 B^A B_A, \\
\mathcal{T}\idx{^{*B}_{BA}}\mathcal{T}\idx{^{*C}_C^A} &= \mathcal{T}\idx{^{B}_{BA}}\mathcal{T}\idx{^{C}_C^A} +6 B_A T^{CA}_C + 9 B^A B_A.
\end{align}
The $BT$ terms are the only possible origin for mixing terms
containing the $B$-field after linearization, and so there will be no
mixing terms in the $a$-matrices if these terms vanish, for which the
condition on the Lagrangian parameters is
\begin{equation}
t_3=\lambda. \label{eqn:ReducePGTCond}
\end{equation}
Moreover, the same condition ensures that the $B^2$ terms from
$\mathcal{T}^{*2}$ also vanish. Hence, if $t_3=\lambda$,
the
$\mathcal{R}+\mathcal{R}^2+\mathcal{T}^{*2}$ part of the WGT$^+$
Lagrangian is identical to its PGT$^+$ counterpart with the
replacement $\mathcal{T}^{*}\rightarrow\mathcal{T}$.

The PGT$^+$ critical cases identified as PPCR in \cite{Lin2020} and having
$t_3=\lambda$ are:
\begin{enumerate}
	\item PGT$^+$ with 2 massless d.o.f. and a massive mode: Case 1, 3, 4, 6, and 7 in Table I of \cite{Lin2020}.
	\item PGT$^+$ with only 2 massless d.o.f.: Case 9-13, 17, and
          19 in Table III of \cite{Lin2020}\footnote{We note
              that cases 9, 10, 11 and 13 in \cite{Lin2020} satisfy the original criterion used by Sezgin in
  \cite{Sezgin1980} to be PCR, and are discussed further in \Cref{sec:PCRcases}}.
	\item PGT$^+$ with only massive mode(s): Cases 26-28, 30-36, and 38-40, 55, and 58 in Table V of \cite{Lin2020}. These cases all have 1 massive mode, either $0^-$ or $2^-$.
\end{enumerate}

If the PGT$^+$ part of a WGT$^+$ satisfying $t_3=\lambda$ has no
propagating mode, then the corresponding WGT$^+$ can at most have a
propagating $B$-field. There are 37 critical cases of PGT$^+$
satisfying $t_3=\lambda$ and containing no propagating mode
(these are
not listed in \cite{Lin2019a} and \cite{Lin2020}). Requiring
$\xi\neq0$ in the corresponding WGT$^+$ Lagrangian (\ref{eqn:WGTLag})
ensures that they contain a propagating dilaton.  The dilaton part of
WGT$^+$ Lagrangians satisfying $t_3=\lambda$ is simply
\begin{equation}
\mathcal{L}_B = \xi \mathcal{H}^{AB} \mathcal{H}_{AB},
\end{equation}
which is that of a massless $1^-$ vector.

For all cases for which the $a$-matrices are non-mixing, there are no
cross terms of $B$ and the other fields and so adding a mass term for
$B$ in the Lagrangian does not affect the other fields. Hence, if one
adds the term $\frac{1}{2}\nu\De^*_A\phi \De^{*A}\phi$ to such a case,
the only effect is either to make an already propagating $B$-field
massive, or to add a non-propagating $B$-field.  In the former (and
more interesting) case, the corresponding dilaton Lagrangian is a
Proca theory in the Einstein gauge ($\phi_0 = 1$)
\begin{equation}
\mathcal{L}_B = \xi \mathcal{H}^{AB} \mathcal{H}_{AB} + 
\tfrac{1}{2}\nu B_\mu B^\mu,
\end{equation}
and the corresponding no-ghost-and-tachyon condition is $\xi<0$ and
$\nu>0$. With these extensions, one can thus construct more tachyon
and ghost free and PPCR cases for WGT$^+$ from the PGT$^+$ cases with
$t_3=\lambda$.

There are, however, some PPCR critical cases of WGT$^+$ that cannot be
constructed directly from PGT$^+$ in the manner described above.
These cases have non-mixing $b$-matrices, but their $a$-matrices
contain mixing terms. In particular, this occurs when there are mixing
terms $\sim BA$ in the linearized Lagrangian. Since the $B$-field can
be fixed using the additional gauge invariance of the
  critical case, there are
no $BA$ terms in the $b$-matrices.  We list these further PPCR critical
cases in \Cref{tab:WGTUnitaryAndPCMore1,tab:WGTUnitaryAndPCMore2}.
Note that none of these cases is PCR.


\bgroup \def\arraystretch{1.5}
\setlength\tabcolsep{0.2cm}
\begin{longtable*}[e]{@{\extracolsep{\fill}}rlll@{}}
	\caption{Parameter conditions for the PPCR
          critical cases that are ghost and tachyon free and cannot be
          constructed directly from PGT. The parameters listed in
          ``Additional conditions'' must be non-zero to prevent the
          theory becoming a different critical case.}
	\label{tab:WGTUnitaryAndPCMore1} \\*
	\toprule
	\#&Critical condition&Additional conditions&No-ghost-and-tachyon condition\\*
	\colrule
	\noalign{\vspace{3pt}}%
	\endfirsthead
	\noalign{\nobreak\vspace{3pt}}%
	\botrule
	\endlastfoot
	
	\rownumber & \makecell[cl]{$r_1,\frac{r_3}{2}-r_4,t_1,\lambda =0$} &\makecell[cl]{$r_2,r_3,2 r_3+r_5,r_3+2 r_5,t_2,t_3$} & $t_2>0, r_2<0, r_3 \left(2 r_3+r_5\right) \left(r_3+2 r_5\right)<0$  \\
	\rownumber & \makecell[cl]{$r_2,r_1-r_3,r_4,t_1,t_2,\lambda =0$} &\makecell[cl]{$r_1,r_1+r_5,2 r_1+r_5,t_3$} & $r_1 \left(r_1+r_5\right) \left(2 r_1+r_5\right)<0$  \\
	\rownumber & \makecell[cl]{$r_1,r_2,\frac{r_3}{2}-r_4,t_1,t_2,\lambda =0$} &\makecell[cl]{$r_3,2 r_3+r_5,r_3+2 r_5,t_3$} & $r_3 \left(2 r_3+r_5\right) \left(r_3+2 r_5\right)<0$  \\
	\rownumber & \makecell[cl]{$r_1,\frac{r_3}{2}-r_4,t_1,t_2,\lambda =0$} &\makecell[cl]{$r_2,r_3,2 r_3+r_5,r_3+2 r_5,t_3$} & $r_3 \left(2 r_3+r_5\right) \left(r_3+2 r_5\right)<0$  \\
	\rownumber & \makecell[cl]{$r_1,r_2,\frac{r_3}{2}-r_4,t_1,\lambda =0$} &\makecell[cl]{$r_3,2 r_3+r_5,r_3+2 r_5,t_2,t_3$} & $r_3 \left(2 r_3+r_5\right) \left(r_3+2 r_5\right)<0$  \\
	\rownumber & \makecell[cl]{$r_1,r_3,r_4,r_5,\lambda =0$} &\makecell[cl]{$r_2,t_1,t_2,t_1+t_2,t_3$} & $t_2>0, r_2<0$  \\
	\rownumber & \makecell[cl]{$r_1,r_3,r_4,r_5,t_1+t_2,\lambda =0$} &\makecell[cl]{$r_2,t_1,t_3$} & $r_2<0, t_1<0$  \\
	\rownumber & \makecell[cl]{$r_2,r_1-r_3,r_4,r_1+r_5,t_1+t_2,\lambda =0$} &\makecell[cl]{$r_1,t_1,t_3$} & $t_1>0, r_1<0$  \\
	\rownumber & \makecell[cl]{$r_1,r_3,r_4,r_5,t_1,\lambda =0$} &\makecell[cl]{$r_2,t_2,t_3$} & $t_2>0, r_2<0$  \\
	\rownumber & \makecell[cl]{$r_1,r_3,r_4,t_1,\lambda =0$} &\makecell[cl]{$r_2,r_5,t_2,t_3$} & $t_2>0, r_2<0$  \\
	\rownumber & \makecell[cl]{$r_1-r_3,r_4,2 r_1+r_5,t_1,\lambda =0$} &\makecell[cl]{$r_1,r_2,t_2,t_3$} & $t_2>0, r_2<0$  \\
	\rownumber & \makecell[cl]{$r_1,\frac{r_3}{2}-r_4,2 r_3+r_5,t_1,\lambda =0$} &\makecell[cl]{$r_2,r_3,t_2,t_3$} & $t_2>0, r_2<0$  \\
	\rownumber & \makecell[cl]{$r_1,\frac{r_3}{2}-r_4,\frac{r_3}{2}+r_5,t_1,\lambda =0$} &\makecell[cl]{$r_2,r_3,t_2,t_3$} & $t_2>0, r_2<0$  \\

\end{longtable*}
\egroup
\medskip

\setcounter{magicrownumbers}{0}


\bgroup \def\arraystretch{1.5}
\setlength\tabcolsep{0.2cm}
\begin{longtable*}[e]{@{\extracolsep{\fill}}rHccl}
	\caption{Particle content of the
		PPCR critical cases that are
		ghost and tachyon free and cannot be constructed directly from PGT. The column ``$b$
		sectors'' describes the diagonal elements in the
		$b^{-1}$-matrix of each spin-parity sector in the
		sequence $\{0^-,0^+,1^-,1^+,2^-,2^+\}$. Here it is notated as
		$\varphi^n_{v}$ or $\varphi^n_{l}$, where $\varphi$
		is the field, $-n$ is the power of $k$ in the
		element in the $b^{-1}$-matrix when $k$ goes to infinity, $v$ means massive
		pole, and $l$ means massless pole. If $n=\infty$, it
		represents that the diagonal element is zero. If
		$n\leq0$, the field is not propagating. The ``$|$''
		notation denotes the different form of the elements
		of the $b^{-1}$-matrices in different choices of
		gauge fixing, and the ``$\&$'' connects the diagonal
		elements in the same $b^{-1}$-matrix. The
		superscript ``N'' represents that there is non-zero
		off-diagonal term in the $b^{-1}$-matrix.	
}
	\label{tab:WGTUnitaryAndPCMore2} \\
	\toprule
	\#&Critical condition&\makecell[cl]{Massless\\ mode d.o.f.}&\makecell[cl]{Massive \\mode}&$b$ sectors\\*
	\colrule
	\noalign{\vspace{3pt}}%
	\endfirsthead
	
	\multicolumn{5}{l}{TABLE~\ref{tab:WGTUnitaryAndPCMore2} (continued)}
	\rule{0pt}{12pt}\\
	\noalign{\vspace{1.5pt}}
	\colrule\rule{0pt}{12pt}
	\#&Critical condition&\makecell[cl]{Massless\\ mode d.o.f.}&\makecell[cl]{Massive \\mode}&$b$ sectors\\*
	\colrule
	\noalign{\vspace{3pt}}%
	\endhead
	\noalign{\nobreak\vspace{3pt}}%
	\colrule
	\endfoot
	\noalign{\nobreak\vspace{3pt}}%
	\botrule
	\endlastfoot
	
	\rownumber & \makecell[cl]{$r_1,\frac{r_3}{2}-r_4,t_1,\lambda =0$} &2 & $0^-$ & $\left\{A\text{}_{\text{v}}^{2},A\text{}_{\text{}}^{0}|\mathfrak{s}\text{}_{\text{l}}^{2}|B\text{}_{\text{}}^{0},\left(A\text{}_{\text{l}}^{2}\&A\text{}_{\text{l}}^{0}\right)^\text{N}|\left(A\text{}_{\text{l}}^{2}\&\mathfrak{s}\text{}_{\text{l}}^{2}\right)^\text{N}|\left(A\text{}_{\text{l}}^{2}\&\mathfrak{a}\text{}_{\text{l}}^{2}\right)^\text{N}|\left(A\text{}_{\text{l}}^{2}\&B\text{}_{\text{l}}^{0}\right)^\text{N},\left(A\text{}_{\text{l}}^{2}\&A\text{}_{\text{l}}^{0}\right)^\text{N}|\left(A\text{}_{\text{l}}^{2}\&\mathfrak{a}\text{}_{\text{l}}^{2}\right)^\text{N},\times,A\text{}_{\text{l}}^{2}\right\}$ \\
	\rownumber & \makecell[cl]{$r_2,r_1-r_3,r_4,t_1,t_2,\lambda =0$} &2 & $\times$ & $\left\{\times,A\text{}_{\text{}}^{0}|\mathfrak{s}\text{}_{\text{l}}^{2}|B\text{}_{\text{}}^{0},\left(A\text{}_{\text{l}}^{2}\&A\text{}_{\text{l}}^{0}\right)^\text{N}|\left(A\text{}_{\text{l}}^{2}\&\mathfrak{s}\text{}_{\text{l}}^{2}\right)^\text{N}|\left(A\text{}_{\text{l}}^{2}\&\mathfrak{a}\text{}_{\text{l}}^{2}\right)^\text{N}|\left(A\text{}_{\text{l}}^{2}\&B\text{}_{\text{l}}^{0}\right)^\text{N},A\text{}_{\text{l}}^{2},A\text{}_{\text{l}}^{2},\times\right\}$ \\
	\rownumber & \makecell[cl]{$r_1,r_2,\frac{r_3}{2}-r_4,t_1,t_2,\lambda =0$} &2 & $\times$ & $\left\{\times,A\text{}_{\text{}}^{0}|\mathfrak{s}\text{}_{\text{l}}^{2}|B\text{}_{\text{}}^{0},\left(A\text{}_{\text{l}}^{2}\&A\text{}_{\text{l}}^{0}\right)^\text{N}|\left(A\text{}_{\text{l}}^{2}\&\mathfrak{s}\text{}_{\text{l}}^{2}\right)^\text{N}|\left(A\text{}_{\text{l}}^{2}\&\mathfrak{a}\text{}_{\text{l}}^{2}\right)^\text{N}|\left(A\text{}_{\text{l}}^{2}\&B\text{}_{\text{l}}^{0}\right)^\text{N},A\text{}_{\text{l}}^{2},\times,A\text{}_{\text{l}}^{2}\right\}$ \\
	\rownumber & \makecell[cl]{$r_1,\frac{r_3}{2}-r_4,t_1,t_2,\lambda =0$} &2 & $\times$ & $\left\{A\text{}_{\text{l}}^{2},A\text{}_{\text{}}^{0}|\mathfrak{s}\text{}_{\text{l}}^{2}|B\text{}_{\text{}}^{0},\left(A\text{}_{\text{l}}^{2}\&A\text{}_{\text{l}}^{0}\right)^\text{N}|\left(A\text{}_{\text{l}}^{2}\&\mathfrak{s}\text{}_{\text{l}}^{2}\right)^\text{N}|\left(A\text{}_{\text{l}}^{2}\&\mathfrak{a}\text{}_{\text{l}}^{2}\right)^\text{N}|\left(A\text{}_{\text{l}}^{2}\&B\text{}_{\text{l}}^{0}\right)^\text{N},A\text{}_{\text{l}}^{2},\times,A\text{}_{\text{l}}^{2}\right\}$ \\
	\rownumber & \makecell[cl]{$r_1,r_2,\frac{r_3}{2}-r_4,t_1,\lambda =0$} &2 & $\times$ & $\left\{A\text{}_{\text{}}^{0},A\text{}_{\text{}}^{0}|\mathfrak{s}\text{}_{\text{l}}^{2}|B\text{}_{\text{}}^{0},\left(A\text{}_{\text{l}}^{2}\&A\text{}_{\text{l}}^{0}\right)^\text{N}|\left(A\text{}_{\text{l}}^{2}\&\mathfrak{s}\text{}_{\text{l}}^{2}\right)^\text{N}|\left(A\text{}_{\text{l}}^{2}\&\mathfrak{a}\text{}_{\text{l}}^{2}\right)^\text{N}|\left(A\text{}_{\text{l}}^{2}\&B\text{}_{\text{l}}^{0}\right)^\text{N},\left(A\text{}_{\text{l}}^{2}\&A\text{}_{\text{l}}^{0}\right)^\text{N}|\left(A\text{}_{\text{l}}^{2}\&\mathfrak{a}\text{}_{\text{l}}^{2}\right)^\text{N},\times,A\text{}_{\text{l}}^{2}\right\}$ \\
	\rownumber & \makecell[cl]{$r_1,r_3,r_4,r_5,\lambda =0$} &0 & $0^-$ & \makecell[cl]{$\left\{A\text{}_{\text{v}}^{2},A\text{}_{\text{}}^{0}|\mathfrak{s}\text{}_{\text{l}}^{2}|B\text{}_{\text{}}^{0},\left(A\text{}_{\text{}}^{0}\&A\text{}_{\text{}}^{0}\right)^\text{N}|\left(A\text{}_{\text{}}^{0}\&\mathfrak{s}\text{}_{\text{l}}^{2}\right)^\text{N}|\left(A\text{}_{\text{}}^{0}\&\mathfrak{a}\text{}_{\text{l}}^{2}\right)^\text{N}|\left(A\text{}_{\text{}}^{0}\&B\text{}_{\text{}}^{0}\right)^\text{N}|\left(\mathfrak{s}\text{}_{\text{l}}^{2}\&B\text{}_{\text{}}^{0}\right)^\text{N}|\left(\mathfrak{a}\text{}_{\text{l}}^{2}\&B\text{}_{\text{}}^{0}\right)^\text{N},\right.$\\$\qquad\left.\left(A\text{}_{\text{}}^{0}\&A\text{}_{\text{}}^{0}\right)^\text{N}|\left(A\text{}_{\text{}}^{0}\&\mathfrak{a}\text{}_{\text{l}}^{2}\right)^\text{N},A\text{}_{\text{}}^{0},A\text{}_{\text{}}^{0}|\mathfrak{s}\text{}_{\text{l}}^{2}\right\}$} \\
	\rownumber & \makecell[cl]{$r_1,r_3,r_4,r_5,t_1+t_2,\lambda =0$} &0 & $0^-$ & \makecell[cl]{$\left\{A\text{}_{\text{v}}^{2},A\text{}_{\text{}}^{0}|\mathfrak{s}\text{}_{\text{l}}^{2}|B\text{}_{\text{}}^{0},\left(A\text{}_{\text{}}^{0}\&A\text{}_{\text{}}^{0}\right)^\text{N}|\left(A\text{}_{\text{}}^{0}\&\mathfrak{s}\text{}_{\text{l}}^{2}\right)^\text{N}|\left(A\text{}_{\text{}}^{0}\&\mathfrak{a}\text{}_{\text{l}}^{2}\right)^\text{N}|\left(A\text{}_{\text{}}^{0}\&B\text{}_{\text{}}^{0}\right)^\text{N}|\left(\mathfrak{s}\text{}_{\text{l}}^{2}\&B\text{}_{\text{}}^{0}\right)^\text{N}|\left(\mathfrak{a}\text{}_{\text{l}}^{2}\&B\text{}_{\text{}}^{0}\right)^\text{N},\right.$\\$\qquad\left.\left(A\text{}_{\text{}}^{\infty}\&A\text{}_{\text{}}^{0}\right)^\text{N}|\left(A\text{}_{\text{}}^{\infty}\&\mathfrak{a}\text{}_{\text{l}}^{2}\right)^\text{N},A\text{}_{\text{}}^{0},A\text{}_{\text{}}^{0}|\mathfrak{s}\text{}_{\text{l}}^{2}\right\}$} \\
	\rownumber & \makecell[cl]{$r_2,r_1-r_3,r_4,r_1+r_5,t_1+t_2,\lambda =0$} &0 & $2^-$ & \makecell[cl]{$\left\{A\text{}_{\text{}}^{0},A\text{}_{\text{}}^{0}|\mathfrak{s}\text{}_{\text{l}}^{2}|B\text{}_{\text{}}^{0},\left(A\text{}_{\text{}}^{0}\&A\text{}_{\text{}}^{0}\right)^\text{N}|\left(A\text{}_{\text{}}^{0}\&\mathfrak{s}\text{}_{\text{l}}^{2}\right)^\text{N}|\left(A\text{}_{\text{}}^{0}\&\mathfrak{a}\text{}_{\text{l}}^{2}\right)^\text{N}|\left(A\text{}_{\text{}}^{0}\&B\text{}_{\text{}}^{0}\right)^\text{N}|\left(\mathfrak{s}\text{}_{\text{l}}^{2}\&B\text{}_{\text{}}^{0}\right)^\text{N}|\left(\mathfrak{a}\text{}_{\text{l}}^{2}\&B\text{}_{\text{}}^{0}\right)^\text{N},\right.$\\$\qquad\left.\left(A\text{}_{\text{}}^{\infty}\&A\text{}_{\text{}}^{-2}\right)^\text{N}|\left(A\text{}_{\text{}}^{\infty}\&\mathfrak{a}\text{}_{\text{l}}^{0}\right)^\text{N},A\text{}_{\text{v}}^{2},A\text{}_{\text{}}^{0}|\mathfrak{s}\text{}_{\text{l}}^{2}\right\}$} \\
	\rownumber & \makecell[cl]{$r_1,r_3,r_4,r_5,t_1,\lambda =0$} &0 & $0^-$ & $\left\{A\text{}_{\text{v}}^{2},A\text{}_{\text{}}^{0}|\mathfrak{s}\text{}_{\text{l}}^{2}|B\text{}_{\text{}}^{0},A\text{}_{\text{}}^{0}|\mathfrak{s}\text{}_{\text{l}}^{2}|\mathfrak{a}\text{}_{\text{l}}^{2}|B\text{}_{\text{}}^{0},A\text{}_{\text{}}^{0}|\mathfrak{a}\text{}_{\text{l}}^{2},\times,\times\right\}$ \\
	\rownumber & \makecell[cl]{$r_1,r_3,r_4,t_1,\lambda =0$} &0 & $0^-$ & $\left\{A\text{}_{\text{v}}^{2},A\text{}_{\text{}}^{0}|\mathfrak{s}\text{}_{\text{l}}^{2}|B\text{}_{\text{}}^{0},\left(A\text{}_{\text{l}}^{2}\&A\text{}_{\text{l}}^{0}\right)^\text{N}|\left(A\text{}_{\text{l}}^{2}\&\mathfrak{s}\text{}_{\text{l}}^{2}\right)^\text{N}|\left(A\text{}_{\text{l}}^{2}\&\mathfrak{a}\text{}_{\text{l}}^{2}\right)^\text{N}|\left(A\text{}_{\text{l}}^{2}\&B\text{}_{\text{l}}^{0}\right)^\text{N},\left(A\text{}_{\text{l}}^{2}\&A\text{}_{\text{l}}^{0}\right)^\text{N}|\left(A\text{}_{\text{l}}^{2}\&\mathfrak{a}\text{}_{\text{l}}^{2}\right)^\text{N},\times,\times\right\}$ \\
	\rownumber & \makecell[cl]{$r_1-r_3,r_4,2 r_1+r_5,t_1,\lambda =0$} &0 & $0^-$ & $\left\{A\text{}_{\text{v}}^{2},A\text{}_{\text{}}^{0}|\mathfrak{s}\text{}_{\text{l}}^{2}|B\text{}_{\text{}}^{0},\left(A\text{}_{\text{l}}^{2}\&A\text{}_{\text{l}}^{0}\right)^\text{N}|\left(A\text{}_{\text{l}}^{2}\&\mathfrak{s}\text{}_{\text{l}}^{2}\right)^\text{N}|\left(A\text{}_{\text{l}}^{2}\&\mathfrak{a}\text{}_{\text{l}}^{2}\right)^\text{N}|\left(A\text{}_{\text{l}}^{2}\&B\text{}_{\text{l}}^{0}\right)^\text{N},A\text{}_{\text{}}^{0}|\mathfrak{a}\text{}_{\text{l}}^{2},A\text{}_{\text{l}}^{2},\times\right\}$ \\
	\rownumber & \makecell[cl]{$r_1,\frac{r_3}{2}-r_4,2 r_3+r_5,t_1,\lambda =0$} &0 & $0^-$ & $\left\{A\text{}_{\text{v}}^{2},A\text{}_{\text{}}^{0}|\mathfrak{s}\text{}_{\text{l}}^{2}|B\text{}_{\text{}}^{0},\left(A\text{}_{\text{l}}^{2}\&A\text{}_{\text{l}}^{0}\right)^\text{N}|\left(A\text{}_{\text{l}}^{2}\&\mathfrak{s}\text{}_{\text{l}}^{2}\right)^\text{N}|\left(A\text{}_{\text{l}}^{2}\&\mathfrak{a}\text{}_{\text{l}}^{2}\right)^\text{N}|\left(A\text{}_{\text{l}}^{2}\&B\text{}_{\text{l}}^{0}\right)^\text{N},A\text{}_{\text{}}^{0}|\mathfrak{a}\text{}_{\text{l}}^{2},\times,A\text{}_{\text{l}}^{2}\right\}$ \\
	\rownumber & \makecell[cl]{$r_1,\frac{r_3}{2}-r_4,\frac{r_3}{2}+r_5,t_1,\lambda =0$} &0 & $0^-$ & $\left\{A\text{}_{\text{v}}^{2},A\text{}_{\text{}}^{0}|\mathfrak{s}\text{}_{\text{l}}^{2}|B\text{}_{\text{}}^{0},A\text{}_{\text{}}^{0}|\mathfrak{s}\text{}_{\text{l}}^{2}|\mathfrak{a}\text{}_{\text{l}}^{2}|B\text{}_{\text{}}^{0},\left(A\text{}_{\text{l}}^{2}\&A\text{}_{\text{l}}^{0}\right)^\text{N}|\left(A\text{}_{\text{l}}^{2}\&\mathfrak{a}\text{}_{\text{l}}^{2}\right)^\text{N},\times,A\text{}_{\text{l}}^{2}\right\}$ \\

\end{longtable*}
\egroup
\setcounter{magicrownumbers}{0}

\phantom{If we set torsion $\mathcal{T}\idx{^{*\rho}_{\mu\nu}}$ to zero, then
from \eqref{eqn:WGTADeltaK} the gauge fields $A\idx{^{AB}_\mu}$ and
$h\idx{_a^\mu}$ are no longer independent. By substituting $A$-fields
by $B$- and $b$-fields, we can apply the same method as in the
previous section to check the no-ghost-and-tachyon conditions of
torsion-free WGT$^+$ and its critical cases. Setting torsion to zero,
the $a$-matrices of the the root theory Lagrangian \eqref{eqn:WGTLag}
are}

\phantom{If we set torsion $\mathcal{T}\idx{^{*\rho}_{\mu\nu}}$ to zero, then
from \eqref{eqn:WGTADeltaK} the gauge fields $A\idx{^{AB}_\mu}$ and
$h\idx{_a^\mu}$ are no longer independent. By substituting $A$-fields
by $B$- and $b$-fields, we can apply the same method as in the
previous section to check the no-ghost-and-tachyon conditions of
torsion-free WGT$^+$ and its critical cases. Setting torsion to zero,
the $a$-matrices of the the root theory Lagrangian \eqref{eqn:WGTLag}
are}

\section{Torsion-free WGT$^+$ \label{sec:WGTTorsionFree}}

As well as the general case of WGT$^+$, one may also consider the
simpler cases with vanishing torsion or curvature, respectively, which
are {\em not} merely special cases of the general WGT$^+$ action,
because additional constraints are placed not only the coefficients,
but also on the fields. In this section we consider the case of
vanishing torsion.


If one sets the torsion $\mathcal{T}\idx{^{*\rho}_{\mu\nu}}$ to zero,
then one sees from \eqref{eqn:WGTADeltaK} that the gauge fields
$A\idx{^{AB}_\mu}$, $h\idx{_a^\mu}$ and $B_\mu$ are no longer
independent. Indeed, \eqref{eqn:WGTADeltaK} gives an explicit
expression for the $A$-field in terms of the $B$- and $b$-fields. On
making this substitution in the Lagrangian, one may then can
apply the same method as in the previous section to investigate
torsion-free WGT$^+$ and its critical cases. In this simpler theory,
one need not set $\nu=\xi=c_1=0$, since one does not encounter
critical conditions that are non-linear in the Lagrangian
parameters. Hence, we do not adopt this restriction in this section.

\subsection{The `root' theory}

In this case, the $a$-matrices of the root theory \eqref{eqn:WGTLag} are
\begin{align}
a(0^+)=&\bordermatrix{
	~&\mathfrak{s} &\mathfrak{s} &B \cr
	\mathfrak{s} & \makecell[cl]{8 \left(r_1-r_3\right.\\\left.+2 r_4\right)k^4 -4\lambda k^2} & 0 &\makecell[cl]{8 i \sqrt{3} \left(r_1-r_3\right.\\\left.+2 r_4\right) k^3}  \cr
	\mathfrak{s} & 0 & 0 & 0 \cr
	B & \makecell[cl]{-8 i \sqrt{3} \left(r_1-r_3\right.\\\left.+2 r_4\right) k^3}  & 0 & \makecell[cl]{24 k^2 \left(r_1-r_3\right.\\\left.+2 r_4\right)+12 \lambda +\nu} \cr
}, \\
a(1^-)=&\bordermatrix{
	~&\mathfrak{s} &\mathfrak{a} &B \cr
	\mathfrak{s}& 0 & 0 & 0 \cr
	\mathfrak{a}& 0 & 0 & 0 \cr
	B& 0 & 0 &\makecell[cl]{4 k^2 \left(c_1+2 r_1+2 r_4+2 r_5+\xi \right)\\+12 \lambda +\nu}\cr
}, \\
a(1^+)=&\bordermatrix{
	~&\mathfrak{a} \cr
	\mathfrak{a}&0 \cr
}, \\
a(2^+)=&\bordermatrix{
	~&\mathfrak{s} \cr
	\mathfrak{s}&  4 \left(2 r_1-2 r_3+r_4\right) k^4 + 2 \lambda k^2 \cr
},
\end{align}
where the SPOs are obtained from those listed in
\Cref{sec:SpinProjectionOperatorWGT} by simply deleting the rows and
columns corresponding to the $A$-field. The $a$-matrices for $0^-$ and
$2^-$ sectors have no element, so we do not list them. One can fix the
gauge simply by removing the rows and columns whose elements are all
zeros from the $a$-matrices, to obtain the corresponding
$b$-matrices. These may then be inverted to obtain the saturated
propagator.

Considering first the massless sector, the nonzero eigenvalues of
the Laurent series coefficient matrix $\mathbf{Q}_{2}$ are:
\begin{equation}
\frac{1}{\lambda},\frac{1}{2\lambda}.
\end{equation}
Thus, the theory has two massless d.o.f., and the no-ghost condition
for the massless sector is simply
\begin{equation}
\lambda>0.
\label{eqn:tfwgtml}
\end{equation}

Turning to the massive sector, the determinants of the $b$-matrices
are
\begin{align}
\mathrm{det}\left[b\left(0^+\right)\right]=&8 \left(r_1-r_3+2 r_4\right) \nu  k^4-4 \lambda  (12 \lambda +\nu )  k^2,\\
\mathrm{det}\left[b\left(1^-\right)\right]=&4 \left(c_1+2 r_1+2 r_4+2 r_5+\xi \right)  k^2+12 \lambda +\nu,\\
\mathrm{det}\left[b\left(2^+\right)\right]=&4 \left(2 r_1-2 r_3+r_4\right)  k^4+2 \lambda  k^2,
\end{align}
from which one obtains the masses
\begin{align}
m^2\left(0^+\right)&=\frac{\lambda\left(12 \lambda+  \nu\right) }{2 \left(r_1-r_3+2 r_4\right) \nu },\\
m^2\left(1^-\right)&=\frac{-12 \lambda -\nu }{4 \left(c_1+2 r_1+2 r_4+2 r_5+\xi \right)},\\
m^2\left(2^+\right)&=-\frac{\lambda }{2 \left(2 r_1-2 r_3+r_4\right)}.
\label{eqn:tfwgtm2}
\end{align}
The no-tachyon conditions $m^2(J^P)>0$ may then be read off from the
above expressions. In each sector, the masses are distinct, and so one
can again apply Eq. (45) in \cite{Lin2019a} directly to obtain the
massive no-ghost conditions
\begin{align}
&0^+: \frac{1}{4 \lambda }+\frac{6 \lambda ^2}{\left(r_1-r_3+2 r_4\right)
\nu ^2}+\frac{3}{\nu }>0,\\
&1^-: c_1+2 \left(r_1+r_4+r_5\right)+\xi <0,\\
&2^+: \lambda <0.
\end{align}
One thus finds that the combined no-ghost-and-tachyon conditions for
the massive sector are
\begin{align}
&0^+:r_1+2 r_4>r_3, \lambda  \nu  (12 \lambda +\nu )>0,
\\
&1^-:12 \lambda +\nu >0, c_1+2 \left(r_1+r_4+r_5\right)+\xi <0,
\\
&2^+:2 r_1+r_4>2 r_3, \lambda <0.
\end{align}

Since the conditions in the massive $2^+$ sector contradict the
condition (\ref{eqn:tfwgtml}) in the massless sector, the theory must
have a ghost or tachyon.

\subsection{Critical cases}


\begin{figure}[t!]
	\includegraphics[width=0.48\textwidth]{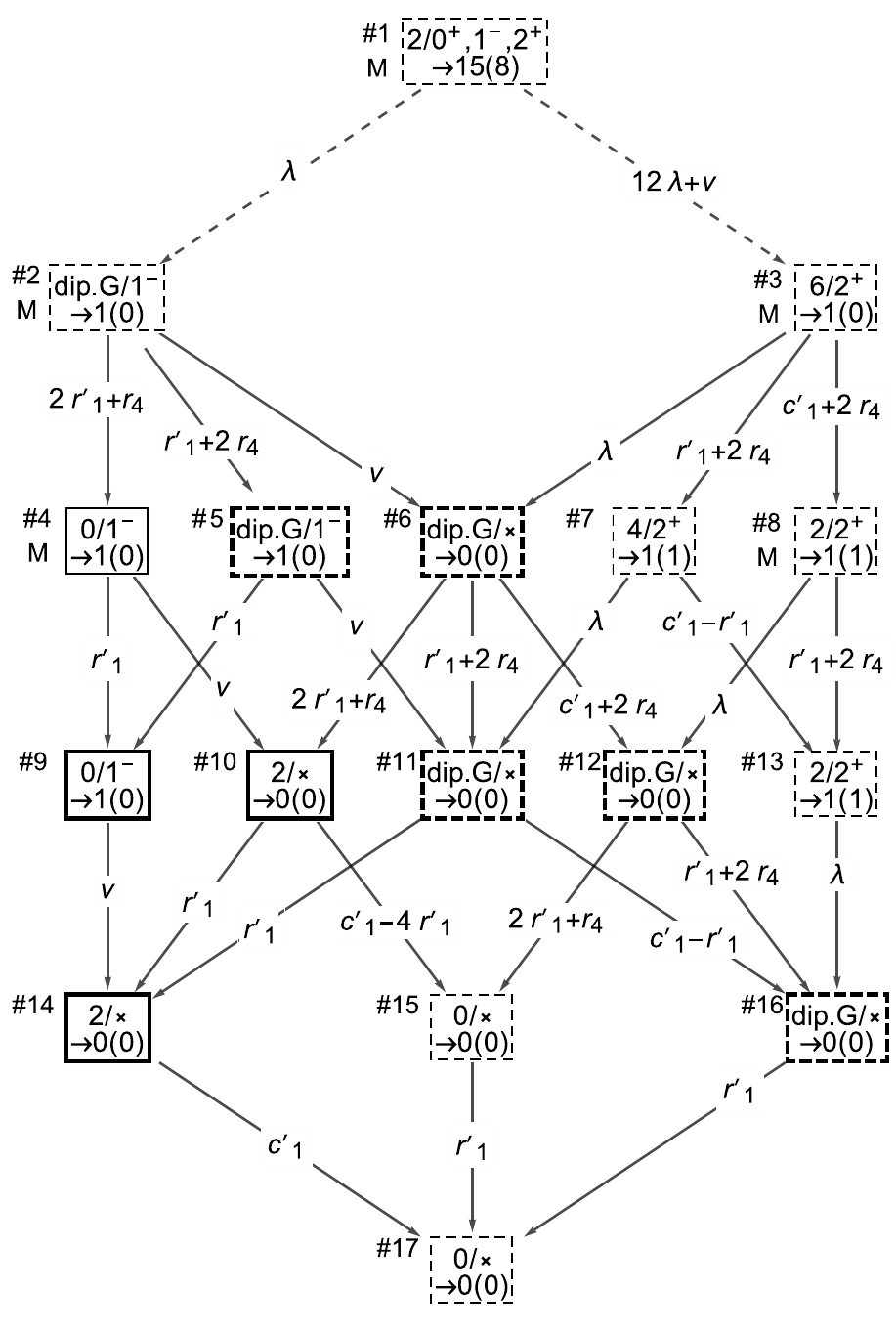}
	\caption{\label{fig:WGTNoTorsionCritical} 
		  Critical cases of torsionless WGT$^+$
          resulting from type A or type B conditions. Each
          node represents a critical case, except the top and bottom
          nodes, which represent the root theory and the zero
          Lagrangian, respectively. Each arrow points from a node to
          one of its critical cases. A solid arrow represents type A
          critical condition, and a dashed arrow represents type B. The
          labels on the arrows are the critical parameters; for
          brevity, the variables $r'_1=r_1-r_3$ and
          $c'_1=c_1+2r_1+2r_5+\xi$ have been defined. The critical
          condition of a node can be obtained by setting all the
          critical parameters to zeros in the path from the root
          theory to that node, and the conditions are path
          independent. In each node, the first line is in the format
          ``d.o.f. of massless mode or `dip.G' if there are massless
          dipole ghosts/massive mode'', and the second line is
          ``number of child critical cases resulting form type C conditions (number of
          no-ghost-and-tachyon cases among them)'', which are not
          shown but are listed in Table~\ref{tab:WGTTorsionless}. The
          dashed/solid frames indicate those cases that contain any/no
          ghost or tachyon. The thick frames indicate PPCR cases, and
          the thin frames indicate those that are non-PPCR or have
          mixing $b$-matrices. The ``M'' under the number at the left
          of the nodes with mixing $b$-matrices.\label{fig:typeab}}
\end{figure}


We now consider the critical cases of torsion-free WGT$^+$. As
discussed in detail in \cite{Lin2019a}, one finds all conditions that
cause a theory to be a critical case. While some conditions may cause
criticality in more than one way, one can still divide all the
critical conditions into three categories, which we called type A, B
and C conditions, respectively.

Considering first the root theory, it becomes critical and thereby
loses one d.o.f in the Lagrangian parameter space if any of the
following expressions vanishes:
\begin{align*}
\text{Type B: }&\lambda ,12 \lambda +\nu,\numberthis
\label{eqn:typeb} \\ \text{Type C:
}&2 r_1-2 r_3+r_4,r_1-r_3+2 r_4,\nu ,\\&c_1+\xi +2 r_1+2 r_4+2
r_5.\numberthis
\label{eqn:typec}
\end{align*}
The two critical cases resulting from the type
  B conditions (\ref{eqn:typeb}) of the root theory contain ghosts or
tachyons, but some of their descendant critical cases, all of which
result from type A or C conditions, are free from
ghosts and tachyons.  The critical cases resulting from type
A and type B conditions of torsion-free
WGT$^+$ are shown in Figure~\ref{fig:typeab}, whereas those arising from type C critical conditions
are listed in Table~\ref{tab:WGTTorsionless}; 
those cases that are ghost-and-tachyon-free are indicated, as
described in the captions. One sees that four 
cases in Figure~\ref{fig:typeab} are free from
ghosts and tachyons, and nine critical cases
in Table~\ref{tab:WGTTorsionless} share this property. We
also note that there are 15 critical cases of the root theory
in total that result from type C conditions, which correspond to self-consistent combinations of those in
(\ref{eqn:typec}).  As is clear from (\ref{eqn:tfwgtm2}), those
critical cases resulting from type C conditions and for which $2 r_1-2 r_3+r_4=0$ are free from ghosts and
tachyons because the $2^+$ massive mode is not propagating.


\bgroup \def\arraystretch{1.3}
\setlength\tabcolsep{0.2cm}
\setlength\LTcapwidth{0.48\textwidth}
\begin{longtable}{@{\extracolsep{\fill}}lllll}
	\caption{Critical cases of
          torsion-free WGT$^+$ resulting from type C conditions. The first
          numbers in the column ``\#'' correspond to the numbers in
          \Cref{fig:WGTNoTorsionCritical}, and the corresponding nodes
          are the parent critical cases of the rows. The ``Critical
          condition'' column indicates the critical condition with
          respect to the parent case. For example, ``\#1-3'' is the
          third critical case resulting from type C conditions of case \#1. The symbols
          `$\circ$'/`$\times$' indicate whether a condition is
          true/false. The ``$-$'' symbols denote that there is no
          propagating mode, and the ``M'' symbols indicate the cases
          with mixing $b$-matrices.}
	\label{tab:WGTTorsionless} \\
	\toprule
	\#&Critical condition&\makecell[cl]{Massive \\mode}&\makecell[cl]{No-ghost\\-and\\-tachyon}&PPCR\\*
	\colrule
	\noalign{\vspace{3pt}}%
	\endfirsthead
	
	\multicolumn{5}{l}{TABLE~\ref{tab:WGTTorsionless} (continued)}
	\rule{0pt}{12pt}\\*
	\noalign{\vspace{1.5pt}}
	\colrule\rule{0pt}{12pt}
	\#&Critical condition&\makecell[cl]{Massive \\mode}&\makecell[cl]{No-ghost\\-and\\-tachyon}&PPCR\\*
	\colrule
	\noalign{\vspace{3pt}}%
	\endhead
	%
	
	\colrule
	\endfoot
	%
	\botrule
	\endlastfoot
	
	\#1-1& \makecell[cl]{$\nu$} &$1^-,2^+$ & $\times$ & $\text{M}$ \\
	\#1-2& \makecell[cl]{$r'_1+2 r_4$} &$1^-,2^+$ & $\times$ & $\times$ \\
	\#1-3& \makecell[cl]{$r'_1+2 r_4,\nu $} &$1^-,2^+$ & $\times$ & $\times$ \\
	\#1-4& \makecell[cl]{$c'_1+2 r_4$} &$0^+,2^+$ & $\times$ & $\text{M}$ \\
	\#1-5& \makecell[cl]{$\nu ,c'_1+2 r_4$} &$2^+$ & $\times$ & $\text{M}$ \\
	\#1-6& \makecell[cl]{$r'_1+2 r_4,c'_1+2 r_4$} &$2^+$ & $\times$ & $\times$ \\
	\#1-7& \makecell[cl]{$r'_1+2 r_4,\nu ,c'_1+2 r_4$} &$2^+$ & $\times$ & $\times$ \\
	\#1-8& \makecell[cl]{$2 r'_1+r_4$} &$0^+,1^-$ & $\circ$ & $\text{M}$ \\
	\#1-9& \makecell[cl]{$2 r'_1+r_4,\nu $} &$1^-$ & $\circ$ & $\text{M}$ \\
	\#1-10& \makecell[cl]{$2 r'_1+r_4,r'_1+2 r_4$} &$1^-$ & $\circ$ & $\times$ \\
	\#1-11& \makecell[cl]{$2 r'_1+r_4,r'_1+2 r_4,\nu $} &$1^-$ & $\circ$ & $\times$ \\
	\#1-12& \makecell[cl]{$2 r'_1+r_4,c'_1+2 r_4$} &$0^+$ & $\circ$ & $\text{M}$ \\
	\#1-13& \makecell[cl]{$2 r'_1+r_4,\nu ,c'_1+2 r_4$} &$\times$ & $\circ$ & $\text{M}$ \\
	\#1-14& \makecell[cl]{$2 r'_1+r_4,r'_1+2 r_4,$\\$c'_1+2 r_4$} &$\times$ & $\circ$ & $\times$ \\
	\#1-15& \makecell[cl]{$2 r'_1+r_4,r'_1+2 r_4,\nu ,$\\$c'_1+2 r_4$} &$\times$ & $\circ$ & $\times$ \\
	\#2-1& \makecell[cl]{$c'_1+2 r_4$} &$\times$ & $\times$ & $\text{M}$ \\
	\#3-1& \makecell[cl]{$2 r'_1+r_4$} &$\times$ & $\times$ & $\text{M}$ \\
	\#4-1& \makecell[cl]{$c'_1-4 r'_1$} &$\times$ & $-$ & $-$ \\
	\#5-1& \makecell[cl]{$c'_1-r'_1$} &$\times$ & $\times$ & $\circ$ \\
	\#7-1& \makecell[cl]{$r'_1$} &$\times$ & $\circ$ & $\times$ \\
	\#8-1& \makecell[cl]{$2 r'_1+r_4$} &$\times$ & $\circ$ & $\text{M}$ \\
	\#9-1& \makecell[cl]{$c'_1$} &$\times$ & $-$ & $-$ \\
	\#13-1& \makecell[cl]{$r'_1$} &$\times$ & $\circ$ & $\times$ \\
\end{longtable}
\egroup


\subsection{Comparison with previous results}

The particle spectrum of a subset of torsion-free Weyl-invariant
higher-curvature gravity theories has been studied previously by
\cite{Tanhayi2012}, both in (anti-)de Sitter and Minkowski backgrounds
(to our knowledge, this is the only other investigation of a
torsionless WGT ground-state in the literature).
For $n=4$ spacetime
dimensions, the coefficients $(\alpha,\beta,\gamma,\epsilon,\sigma)$
in their Lagrangian (see equations (1), (7) and (14) in
\cite{Tanhayi2012}) are related to those in our notation used in
(\ref{eqn:WGTLag}) by
\begin{align*}
\alpha&=-\tfrac{1}{2}r_1+r_3=\tfrac{1}{4}(r_4-r_5),\\
\beta&=r_4+r_5=-\tfrac{1}{2}c_1,\\
\gamma&=\tfrac{1}{2}r_1,\\
\epsilon&=\xi-(r_4+r_5+2r_1),\\
\sigma &= \lambda,\numberthis
\label{eqn:compare1}
\end{align*}
together with the conditions
\begin{equation}
	r_1=r_2, \qquad \nu=-1.
\label{eqn:compare2}
\end{equation}
In particular, one should note that the Lagrangian in
\cite{Tanhayi2012} is written in terms of the curvature tensor
$\tilde{\mathcal{R}}_{\mu\nu\rho\sigma}$. As discussed in
Section~\ref{sec:wgt}, this has even fewer symmetry properties than
the rotational gauge field strength tensor
$\mathcal{R}_{\mu\nu\rho\sigma}$ used in
(\ref{eqn:WGTLag}). Consequently, there are further quadratic
combinations of $\tilde{\mathcal{R}}_{\mu\nu\rho\sigma}$ that could
appear in the Lagrangian in \cite{Tanhayi2012}, but only three such
terms are included. Consequently, there are fewer degrees of freedom
in the parameters of their Lagrangian, as compared with our Lagrangian
in (\ref{eqn:WGTLag}), as is evident from the above parameter
identifications. Moreover, since
$\tilde{\mathcal{R}}_{\mu\nu\rho\sigma}$ has many fewer symmetries
than the standard curvature tensor in Riemannian spacetime $V_4$, the
appropriate form of the Gauss--Bonnet identity differs from the usual
formula that is assumed in Eq. (34) of \cite{Tanhayi2012} (see, for
example \cite{Oda2020,Lasenby2016}); fortunately most of the
conclusions presented in \cite{Tanhayi2012} do not depend on this
expression.

The constraints on our parameters in
(\ref{eqn:compare1})--(\ref{eqn:compare2}) do not coincide with any of
the critical conditions in any critical case, so the structure of our
`criticality tree' of torsion-free WGT is not affected. In
\cite{Tanhayi2012}, it is found that about a 4-dimensional Minkowski
background, the WGTs considered are unitary provided (in terms of
our parameters)
\begin{align}
2(r_1-r_3)+r_4&=0, \label{eqn:TanhayiCondition1}\\
r_1-r_3+2r_4&=0,\label{eqn:TanhayiCondition2}\\
\lambda>0.
\end{align} 
Both equalities coincide with our type C critical conditions, and they
eliminate $2^+$ and $0^+$ massive modes, leaving a $1^-$ massive
mode. The condition on $\lambda$ also matches ours, so their result is
consistent with our critical case \#1-10 of the root theory,
listed in Table~\ref{tab:WGTTorsionless}.  

It is concluded in \cite{Tanhayi2012}, however, that the theory has a
massless spin-2 field and a massless spin-0 field, and so the massless
sector has 3 d.o.f, whereas we find just 2.  This difference may
result from the fact that they employ a gauge fixing condition
$\mathcal{D}^*_\mu B^\mu=0$ on the $B^\mu$-field (their
$A^\mu$-field), described in their Eq. (30), but then treat 
this field as if it is unconstrained when reading off the particle
content from their Eq. (59). This situation is analogous to that in
Stueckelberg theory, as discussed in Appendix B in \cite{Lin2020}. If
one fixes the gauge by setting $\de\cdot B=0$, then the Lagrangian
appears to describe a massive vector $B$ and a massless scalar $\phi$
without interaction. Conversely, if one instead sets $\phi=0$, the
Lagrangian contains only a massive vector without constraint. Thus,
one should interpret the theory as containing either a massive vector
or a massive vector with a Stueckelberg ghost and a Faddeev--Popov
ghost.

Also, it is claimed in \cite{Tanhayi2012} that unitarity requires both
\eqref{eqn:TanhayiCondition1} and \eqref{eqn:TanhayiCondition2} to
hold, whereas we require only the former condition, if no Type A or B
critical condition is satisfied. The condition
\eqref{eqn:TanhayiCondition2} is necessary in \cite{Tanhayi2012}
because they do not adopt the Einstein gauge, and so require the
higher-derivative Pais--Uhlenbeck term $(\bar{\square}\Phi_L)^2$ to
vanish, where $\Phi_L$ is the linearized $\phi$. By contrast, all the
higher-order poles in our saturated propagator vanish due to the
source constraints, and so the condition \eqref{eqn:TanhayiCondition2}
is not necessary in our case. This difference may be worthy of further
investigation.

\subsection{Propagating power-counting renormalizability}

We determine whether each critical case is PPCR using the same method
as discussed in Section~\ref{sec:PCR}. The results are presented in
Figure~\ref{fig:typeab} and Table~\ref{tab:WGTTorsionless}. In
particular, we find three critical cases in
  Figure~\ref{fig:typeab} that are both PPCR and contain no ghost or
tachyon; these are indicated by
nodes with thick, solid frames. We note that each
  of these theories can be gauge fixed to contain only the $B$ gauge
  field. It is also worth highlighting that, perhaps as a
  consequence of this, there is no
simultaneously unitary and PPCR case in torsion-free
PGT$^+$ \cite{Lin2019a}, and so these three theories may be
  worthy of further investigation. No 
critical case in Table~\ref{tab:WGTTorsionless} is both PPCR
and unitary.

\section{Curvature-free WGT$^+$ \label{sec:WGTCurvatureFree}}
In this section, we consider WGT$^+$ with vanishing
curvature. This is a more subtle condition than the equivalent case in
PGT$^+$, which was discussed in \cite{Lin2019a}.  As mentioned in
Section~\ref{sec:wgt}, the geometric (Riemann) curvature tensor
$\tilde{\mathcal{R}}\idx{^\rho_{\sigma\mu\nu}}$ in Weyl--Cartan
spacetime differs from the rotational gauge field strength
$\mathcal{R}\idx{^\rho_{\sigma\mu\nu}}$, so it is unclear which should
be set to zero. Here we consider only the case in which the latter
vanishes, since this may imposed in the same way as in PGT by simply
setting $A_{AB\mu}=0$, since the expression for the rotational gauge
field strength in terms of the rotational gauge field are identical in
PGT and WGT. In this simpler theory, one sees from (\ref{eqn:WGTLag})
that one requires only the Lagrangian parameters $\xi$, $\nu$, $t_1$, $t_2$ and
$t_3$, since one can set $\lambda = 0$ without loss of generality.

\subsection{The `root' theory}

In this case, the $a$-matrices of the root theory are
\begin{align*}
&a\left(0^+\right)=\bordermatrix{
	\text{}&\mathfrak{s}&\mathfrak{s}&B\cr
	\mathfrak{s}&4 k^2 t_3&0&4 i \sqrt{3} k t_3\cr
	\mathfrak{s}&0&0&0\cr
	B&-4 i \sqrt{3} k t_3&0&12 t_3+\nu\cr
}, \numberthis\\
&a\left(1^-\right)=\\
&\bordermatrix{
	\text{}&\mathfrak{s}&\mathfrak{a}&B\cr
	\mathfrak{s}&\frac{2}{3} k^2 \left(t_1+t_3\right)&-\frac{2}{3} k^2 \left(t_1+t_3\right)&-2 i \sqrt{2} k t_3\cr
	\mathfrak{a}&-\frac{2}{3} k^2 \left(t_1+t_3\right)&\frac{2}{3} k^2 \left(t_1+t_3\right)&2 i \sqrt{2} k t_3\cr
	B&2 i \sqrt{2} k t_3&-2 i \sqrt{2} k t_3&\makecell[cc]{12 t_3+\nu \\+4 k^2 \xi}\cr
}, \numberthis\\
&a\left(1^+\right)=\bordermatrix{
	\text{}&\mathfrak{a}\cr
	\mathfrak{a}&\frac{2}{3} k^2 \left(t_1+t_2\right)\cr
}, \numberthis\\
&a\left(2^+\right)=\bordermatrix{
	\text{}&\mathfrak{s}\cr
	\mathfrak{s}&2 k^2 t_1\cr
}.\numberthis
\end{align*}
As in the torsion-free theory, the SPOs are obtained from those listed
in \Cref{sec:SpinProjectionOperatorWGT} by deleting the rows and
columns corresponding to the $A$-field, and the $a$-matrices for the
$0^-$ and $2^-$ sectors contain no elements. After fixing the gauge by
deleting rows and columns, one obtains the non-singular $b$ matrices,
which may be inverted to obtain saturated propagator.

Considering first the massless sector, one finds that the Laurent
series coefficient matrix $\mathbf{Q}_{4}$ is non-zero in this case, and the
condition for it to vanish is
\begin{equation}
\nu =-\frac{12 t_1 \left(t_1-2 t_2\right) t_3}
{t_1^2-2 t_1 t_2+4 t_1 t_3+t_2 t_3}. \label{eqn:WGTNoCurvatureQ4Cond}
\end{equation}
One further finds that the
Laurent coefficient matrix $\mathbf{Q}_{2}$ cannot be positive
definite and contains eight nonzero eigenvalues, which are too
complicated to be given here. Consequently, the root theory must
contain ghosts in the massless sector.

One can, however, continue to analyze the massive sector. The
determinants of the $b$ matrices are
\begin{align}
\det\left[b\left(0^+\right)\right]&=4 t_3 \nu  k^2, \\
\det\left[b\left(1^-\right)\right]&=\tfrac{2}{3} \left[t_3 \nu +t_1 \left(12 t_3+\nu \right)\right]  k^2  \\&+  \tfrac{8}{3} \left(t_1+t_3\right) \xi  k^4, \\
\det\left[b\left(1^+\right)\right]&=\tfrac{2}{3} \left(t_1+t_2\right)  k^2, \\
\det\left[b\left(2^+\right)\right]&=2 t_1  k^2. 
\end{align}
Only the $1^-$ sector contains a massive mode, with mass
\begin{equation}
m^2\left(1^-\right)=\frac{-12 t_1 t_3-(t_1 +t_3) \nu }{4 \left(t_1+t_3\right) \xi },
\label{eqn:cfcond1}
\end{equation}
and the no-tachyon condition is $m^2\left(1^-\right)>0$. Applying
Eq. (45) in \cite{Lin2019a} directly, in this case the no-ghost
condition is
\begin{align*}
1^-:&\left(t_1+t_3\right) \left[12 t_1 t_3+\left(t_1+t_3\right) \nu \right] \xi  \left\{\left(t_1+t_3\right) \left[12 t_1 t_3\right.\right.\\&\left.\left.+\left(t_1+t_3\right) \nu \right]-72 t_3^2 \xi \right\}<0. \numberthis
\end{align*}
The combined no-ghost-and-tachyon conditions for the massive sector are thus
\begin{equation}
\xi <0,\qquad \nu >-\frac{12 t_1 t_3}{t_1+t_3},
\label{eqn:cfcond3}
\end{equation}
but one should recall that the massless sector always contains a ghost.

\subsection{Critical cases}

The critical cases of the root theory occur when any of
the following expressions vanishes:
\begin{align}
\text{Type A: }& t_1,t_1+t_2,t_3,\nu, \label{eqn:cfcrit1}\\ \text{Type B: }& 12 t_1
t_3+t_1 \nu +t_3 \nu, \\ 
\text{Type C: }& t_1+t_3,\xi.\label{eqn:cfcrit3}
\end{align}
However, since $12 t_1 t_3+t_1 \nu +t_3 \nu$ cannot be factorized into
a linear combination of the parameters, one cannot apply our algorithm
to find all the critical cases directly. We therefore below consider the
critical case $\nu=0$, which removes the kinetic term of the scalar
field $\phi$, as the simplified root theory and instead find its
critical cases. Before turning to these, we note that the massless
sector of this simplified root theory requires $t_1-2t_2=0$ to make
its Laurent series coefficient matrix $\mathbf{Q}_{4}$ vanish, and
thus prevent the presence of dipole ghosts, but in any case the matrix
$\mathbf{Q}_{2}$ has seven nonzero eigenvalues and cannot be made be
positive definite. Therefore, the massless sector must contain a
ghost. The conditions for the massive sector of the simplified root
theory to be ghost and tachyon free may be obtained from
(\ref{eqn:cfcond1})--(\ref{eqn:cfcond3}) by setting $\nu=0$.

\begin{figure}[h!]
	\includegraphics[width=0.48\textwidth]{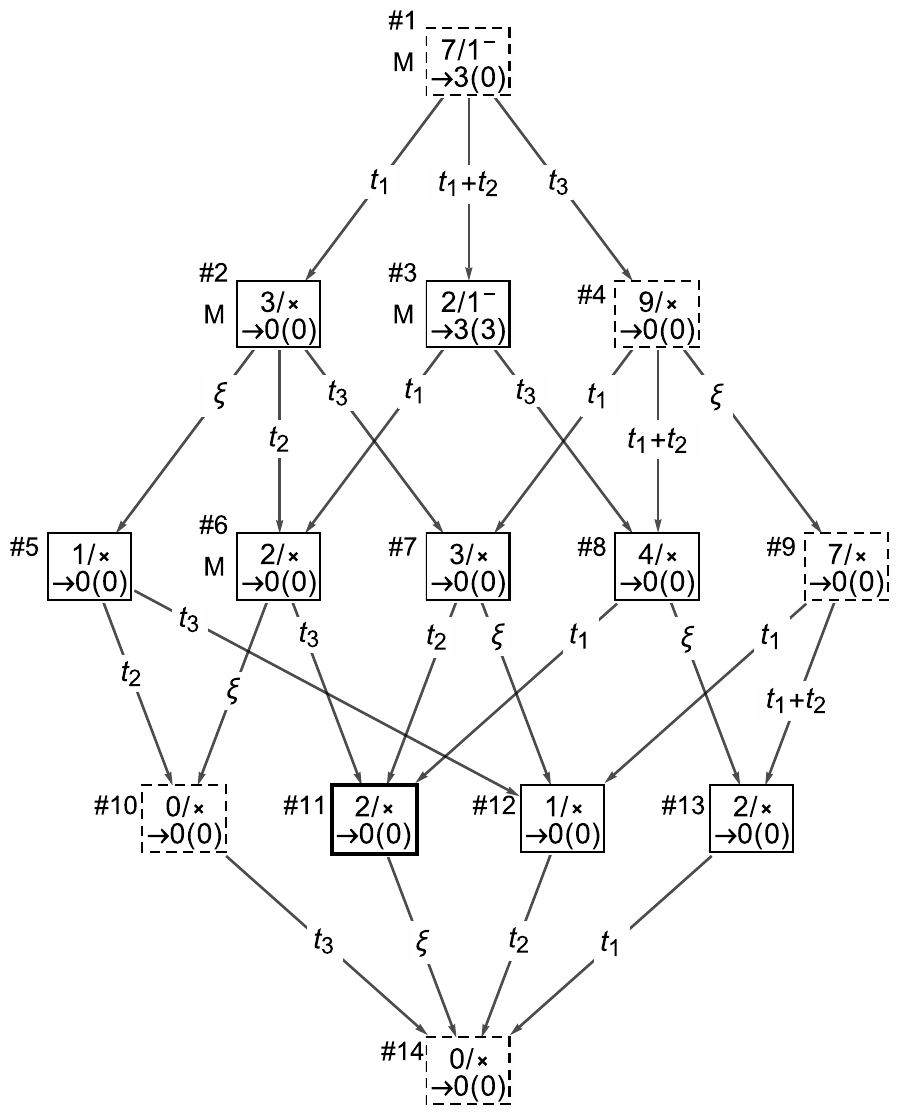}
	\caption{\label{fig:WGTNoCurvatureCritical} Critical cases resulting from type A critical conditions of curvature-free WGT$^+$. The notation follows
          that of \Cref{fig:WGTNoCurvatureCritical}.}
\end{figure}

Turning now to the critical cases of the simplified root theory, the
critical conditions are given by
(\ref{eqn:cfcrit1})--(\ref{eqn:cfcrit3}) with $\nu = 0$. One should
note that this results in the simplified root theory containing no type B critical condition, since the resulting
condition that $t_1t_3$ should vanish is trivially factorised and the
separate requirements that $t_1$ or $t_3$ should vanish are already
included in the type A critical conditions, and it turns out that there is no type B critical condition in the descendants. The critical cases resulting from type A or type C
  conditions are summarised in
Figure~\ref{fig:WGTNoCurvatureCritical} and
Table~\ref{tab:WGTNoCurvature}, respectively. Cases that are
ghost-and-tachyon-free are indicated, as described in the captions. In
particular, we note that there are nine critical cases 
in Figure~\ref{fig:WGTNoCurvatureCritical} that are
free from ghosts and tachyons, and three such critical
cases in Table~\ref{tab:WGTNoCurvature}.
\begin{table}
\caption{Critical cases resulting from type C critical conditions of curvature-free WGT$^+$. The notation
  follows that of \Cref{tab:WGTTorsionless}.\label{tab:WGTNoCurvature}}
\begin{ruledtabular}
\begin{tabular}{lllll}
	\#&Critical condition&\makecell[cl]{Massive \\mode}&\makecell[cl]{No-ghost\\-and\\-tachyon}&PPCR\\
	\colrule
	\#1-1& \makecell[cl]{$\xi$} &$\times$ & $\times$ & $\text{M}$ \\
	\#1-2& \makecell[cl]{$t_1+t_3$} &$\times$ & $\times$ & $\text{M}$ \\
	\#1-3& \makecell[cl]{$t_1+t_3,\xi $} &$\times$ & $\times$ & $\text{M}$ \\
	\#3-1& \makecell[cl]{$\xi$} &$\times$ & $\circ$ & $\text{M}$ \\
	\#3-2& \makecell[cl]{$t_1+t_3$} &$\times$ & $\circ$ & $\text{M}$ \\
	\#3-3& \makecell[cl]{$t_1+t_3,\xi $} &$\times$ & $\circ$ & $\text{M}$ \\
\end{tabular}
\end{ruledtabular}
\end{table}

\subsection{Propagating power-counting renormalizability}

We determine whether each critical case is PPCR using the same method
as discussed in Section~\ref{sec:PCR}. The results are presented in
Figure~\ref{fig:WGTNoCurvatureCritical} and
Table~\ref{tab:WGTNoCurvature}. In particular, we find that there is
just a single critical case in
  Figure~\ref{fig:WGTNoCurvatureCritical}, which is just the pure
dilaton Lagrangian $\mathcal{L}\sim\mathcal{H}^2$, that is both PPCR and unitary; this is indicated
by the node with a
thick, solid frame. No such critical case is found in 
Table~\ref{tab:WGTNoCurvature}.

\section{Conclusions}
We have used the systematic method in \cite{Lin2019a} to determine the
no-ghost-and-tachyon conditions for the most general WGT$^+$ (the root
theory), and found it must contain a ghost or tachyon. For a subset of
the theory, with the restriction $\nu=\xi=c_1=0$ on the parameters in
the Lagrangian (\ref{eqn:WGTLag}), which removes the kinetic terms for
the scalar field $\phi$ and dilational gauge field $B$, respectively,
and the only `cross term' $\mathcal{R}^{AB} \mathcal{H}_{AB}$ between
gauge field strengths, we found and categorised all \numCriticalNotRootNotZero{}
critical cases, and identified \numNoGhostNoTachyon{}
that are free from ghosts and tachyons. The full
set of results displayed in an interactive form can be found at:
\url{http://www.mrao.cam.ac.uk/projects/gtg/wgt/}. We
compared our findings with the only other example of a unitary WGT$^+$
of which we are aware in the literature \cite{Nieh1982}, and found the
results to be consistent. We further identified those critical cases
of WGT$^+$ that are also PPCR. Most of these are identical to those in
PGT$^+$ listed in \cite{Lin2020}, or are a PGT$^+$ without any
propagating mode (which were not listed in
\cite{Lin2020}). Nonetheless, we also identified a further 13 PPCR and
ghost-and-tachyon-free critical cases of WGT$^+$ that cannot be
constructed directly from PGT$^+$.

We repeated our analysis for the simpler cases of torsion-free and
curvature-free WGT$^+$, which are not merely special cases of the
general WGT$^+$ action, because additional constraints are placed not
only the coefficients, but also on the fields.  For the torsion-free
case, we found that the root theory (without any further conditions on
the Lagrangian parameters) must contain a ghost or
tachyon. Nonetheless, we identify 13 critical cases that are free from
ghosts and tachyons. We also compare our results with the only other
invesigation of the ground-state of a torsionless WGT$^+$ of which we
are aware in the literature. We find our results to be consistent,
apart from a minor issue related to the number of propagating degrees
of freedom in the massless sector, most probably resulting from the
different approaches to gauge-fixing used in the two analyses. Of our
13 ghost-and-tachyon-free critical cases, we further identified three
that are also PPCR, each of which can be gauge fixed to contain
  only the $B$ gauge field. This may explain the sharp contrast with
  torsion-free PGT$^+$, for which there is no unitary and PPCR
  critical case, and suggests that these three theories may be worthy
  of further investigation.

For curvature-free WGT$^+$, we find that the massless sector of the
root theory (again with no further conditions on the Lagrangian
parameters) must contain a ghost. For the simplified root theory with
$\nu= 0$, which has no kinetic term for the scalar field $\phi$ in the
Lagrangian and is itself found to have a ghost in the massless sector,
we find 13 critical cases that are free from ghosts and tachyons, of
which just a single case is found also to be PPCR, which
  corresponds to the pure dilaton Lagrangian
  $\mathcal{L}\sim\mathcal{H}^2$.

All the restrictions on Lagrangian parameters mentioned above are
necessary to avoid critical conditions that cannot be written as the
product of real linear terms, which is required by the systematic
method in \cite{Lin2019a}. We plan to improve our approach to
accommodate such cases in future work, and also apply the method to
more general gauge theories, such as metric affine gravities, whose
unitarity was recently investigated by \cite{Percacci2020} using SPOs.

Finally, we point out that gauge theories of gravity can yield interesting
phenomenology. In particular, in a cosmological context, recent
investigations of some of the PGT$^+$ cases that were identified in
\cite{Lin2019a,Lin2020} as being unitary and PPCR have been carried out in
\cite{Barker2020a}, and are found to have rich background solutions that
support the concordance $\Lambda$CDM background cosmology up to an
optional, effective dark radiation, which shows considerable promise in
alleviating the Hubble tension. These theories have been shown to map to a
noncanonical biscalar-tensor theory in the Jordan frame, which provides a
unified framework for future investigation by the broader community, and
for many parameter choices the noncanonical term reduces to a Cuscuton
field \cite{Barker2020b}. Moreover, one of the cases yields two dark energy
solutions: accelerated expansion from a negative bare cosmological constant
whose magnitude is screened, and emergent dark energy to replace vanishing
bare cosmological constant in $\Lambda$CDM. Further investigation of the
unitary and PPCR cases of PGT$^+$ and WGT$^+$ is ongoing.

\begin{acknowledgments}
	Y.-C. Lin acknowledges support from the Ministry of Education
        of Taiwan and the Cambridge Commonwealth, European \&
        International Trust via a Taiwan Cambridge Scholarship and
        financial support from Corpus Christi College, Cambridge.
\end{acknowledgments}

\onecolumngrid
\appendix

\section{Spin Projection Operators for WGT$^+$ \label{sec:SpinProjectionOperatorWGT}}

The block matrices $\mathsf{P}(J^P)$ containing the spin projection
operators for WGT$^+$ used in this paper are as follows:

\begin{align}
&\mathsf{P}\left( 0^-\right)=
\bordermatrix{
	~& A_{ABC} \cr
	A^*_{IJK}& \frac{2}{3} \Theta \text{}_I\text{}_C \Theta \text{}_J\text{}_A \Theta \text{}_K\text{}_B+\frac{1}{3} \Theta \text{}_I\text{}_A \Theta \text{}_J\text{}_B \Theta \text{}_K\text{}_C \cr
},
\\[3mm]
&\mathsf{P}\left( 0^+\right)=
\bordermatrix{
	~&A_{KIJ}&s_{AB}&s_{AB}&B_{C} \cr
	A^*_{CAB}&\frac{2}{3} \Theta \text{}_C\text{}_B \Theta \text{}_K\text{}_J \Omega \text{}_I\text{}_A & \frac{\sqrt{2}}{3} \tilde{k}\text{}_J \Theta \text{}_A\text{}_B \Theta \text{}_K\text{}_I & \sqrt{\frac{2}{3}} \tilde{k}\text{}_J \Theta \text{}_K\text{}_I \Omega \text{}_B\text{}_A & -\sqrt{\frac{2}{3}} \Theta \text{}_K\text{}_J \Omega \text{}_I\text{}_C \cr
	s^*_{IJ}&\frac{\sqrt{2}}{3} \tilde{k}\text{}_B \Theta \text{}_C\text{}_A \Theta \text{}_I\text{}_J & \frac{1}{3} \Theta \text{}_A\text{}_B \Theta \text{}_I\text{}_J & \frac{1}{\sqrt{3}}\Theta \text{}_I\text{}_J \Omega \text{}_A\text{}_B & \frac{1}{\sqrt{3}}\tilde{k}\text{}_C \Theta \text{}_I\text{}_J \cr
	s^*_{IJ}&\sqrt{\frac{2}{3}}\tilde{k}\text{}_B \Theta \text{}_C\text{}_A \Omega \text{}_J\text{}_I & \frac{1}{\sqrt{3}}\Theta \text{}_A\text{}_B \Omega \text{}_I\text{}_J & \Omega \text{}_A\text{}_B \Omega \text{}_I\text{}_J & \tilde{k}\text{}_C \Omega \text{}_I\text{}_J \cr
	B^*_{K}&-\sqrt{\frac{2}{3}} \Theta \text{}_C\text{}_B \Omega \text{}_A\text{}_K & \frac{1}{\sqrt{3}}\tilde{k}\text{}_K \Theta \text{}_A\text{}_B & \tilde{k}\text{}_K \Omega \text{}_A\text{}_B & \Omega \text{}_K\text{}_C \cr
},
\\[3mm]
&\mathsf{P}\left( 1^-\right)=
\bordermatrix{
	~&A_{ABC}&A_{ABC}&s_{AB}&a_{AB}&B_{C} \cr
	A^*_{KIJ}&\Theta \text{}_C\text{}_B \Theta \text{}_I\text{}_A \Theta \text{}_K\text{}_J & \sqrt{2} \Theta \text{}_I\text{}_A \Theta \text{}_K\text{}_J \Omega \text{}_C\text{}_B & \sqrt{2} \tilde{k}\text{}_B \Theta \text{}_I\text{}_A \Theta \text{}_K\text{}_J & \sqrt{2} \tilde{k}\text{}_B \Theta \text{}_I\text{}_A \Theta \text{}_K\text{}_J & \Theta \text{}_I\text{}_C \Theta \text{}_K\text{}_J \cr
	A^*_{KIJ}&\sqrt{2} \Theta \text{}_A\text{}_I \Theta \text{}_C\text{}_B \Omega \text{}_K\text{}_J & 2 \Theta \text{}_I\text{}_A \Omega \text{}_C\text{}_B \Omega \text{}_K\text{}_J & 2 \tilde{k}\text{}_J \Theta \text{}_I\text{}_A \Omega \text{}_K\text{}_B & 2 \tilde{k}\text{}_J \Theta \text{}_I\text{}_A \Omega \text{}_K\text{}_B & \sqrt{2} \Theta \text{}_I\text{}_C \Omega \text{}_K\text{}_J \cr
	s^*_{IJ}&\sqrt{2} \tilde{k}\text{}_J \Theta \text{}_A\text{}_I \Theta \text{}_C\text{}_B & 2 \tilde{k}\text{}_B \Theta \text{}_A\text{}_I \Omega \text{}_C\text{}_J & 2 \Theta \text{}_I\text{}_A \Omega \text{}_J\text{}_B & 2 \Theta \text{}_I\text{}_A \Omega \text{}_J\text{}_B & \sqrt{2} \tilde{k}\text{}_J \Theta \text{}_I\text{}_C \cr
	a^*_{IJ}&\sqrt{2} \tilde{k}\text{}_J \Theta \text{}_A\text{}_I \Theta \text{}_C\text{}_B & 2 \tilde{k}\text{}_B \Theta \text{}_I\text{}_A \Omega \text{}_C\text{}_J & 2 \Theta \text{}_I\text{}_A \Omega \text{}_J\text{}_B & 2 \Theta \text{}_I\text{}_A \Omega \text{}_J\text{}_B & \sqrt{2} \tilde{k}\text{}_J \Theta \text{}_I\text{}_C \cr
	B^*_{K}&\Theta \text{}_A\text{}_K \Theta \text{}_C\text{}_B & \sqrt{2} \Theta \text{}_A\text{}_K \Omega \text{}_C\text{}_B & \sqrt{2} \tilde{k}\text{}_B \Theta \text{}_A\text{}_K & \sqrt{2} \tilde{k}\text{}_B \Theta \text{}_A\text{}_K & \Theta \text{}_K\text{}_C \cr
},
\\[3mm]
&\mathsf{P}\left( 1^+\right)=
\bordermatrix{
	~&A_{ABC}&A_{ABC}&a_{AB} \cr
	A^*_{IJK}&\Theta \text{}_I\text{}_C \Theta \text{}_K\text{}_B \Omega \text{}_J\text{}_A+\Theta \text{}_I\text{}_A \Theta \text{}_K\text{}_C \Omega \text{}_J\text{}_B & -\sqrt{2} \Theta \text{}_J\text{}_A \Theta \text{}_K\text{}_B \Omega \text{}_I\text{}_C & \sqrt{2} \tilde{k}\text{}_J \Theta \text{}_I\text{}_A \Theta \text{}_K\text{}_B \cr
	A^*_{IJK}&-\sqrt{2} \Theta \text{}_B\text{}_I \Theta \text{}_C\text{}_J \Omega \text{}_A\text{}_K & \Theta \text{}_I\text{}_A \Theta \text{}_J\text{}_B \Omega \text{}_K\text{}_C & \tilde{k}\text{}_K \Theta \text{}_I\text{}_A \Theta \text{}_J\text{}_B \cr
	a^*_{IJ}&\sqrt{2} \tilde{k}\text{}_B \Theta \text{}_A\text{}_I \Theta \text{}_C\text{}_J & \tilde{k}\text{}_C \Theta \text{}_A\text{}_I \Theta \text{}_B\text{}_J & \Theta \text{}_A\text{}_I \Theta \text{}_B\text{}_J \cr
},
\\[3mm]
&\mathsf{P}\left( 2^-\right)=
\bordermatrix{
	~&A_{ABC} \cr
	A^*_{IJK}&\frac{2}{3} \Theta \text{}_I\text{}_C \Theta \text{}_J\text{}_B \Theta \text{}_K\text{}_A+\frac{2}{3} \Theta \text{}_I\text{}_A \Theta \text{}_J\text{}_B \Theta \text{}_K\text{}_C-\Theta \text{}_C\text{}_B \Theta \text{}_I\text{}_A \Theta \text{}_K\text{}_J \cr
}, \\[3mm]
&\mathsf{P}\left(2^+\right)=
\bordermatrix{
	~&A_{ABC}&s_{AB} \cr
	A^*_{IJK}&-\frac{2}{3} \Theta \text{}_C\text{}_B \Theta \text{}_K\text{}_J \Omega \text{}_I\text{}_A+\Theta \text{}_I\text{}_C \Theta \text{}_K\text{}_A \Omega \text{}_J\text{}_B+\Theta \text{}_I\text{}_A \Theta \text{}_K\text{}_C \Omega \text{}_J\text{}_B & \sqrt{2} \tilde{k}\text{}_J \left(\Theta \text{}_I\text{}_A \Theta \text{}_K\text{}_B-\frac{1}{3} \Theta \text{}_A\text{}_B \Theta \text{}_K\text{}_I\right) \cr
	s^*_{IJ}&\sqrt{2} \tilde{k}\text{}_B \left(\Theta \text{}_C\text{}_J \Theta \text{}_I\text{}_A-\frac{1}{3} \Theta \text{}_C\text{}_A \Theta \text{}_I\text{}_J\right) & -\frac{1}{3} \Theta \text{}_A\text{}_B \Theta \text{}_I\text{}_J+\Theta \text{}_I\text{}_A \Theta \text{}_J\text{}_B \cr
}.\\\nonumber
\end{align}
These SPOs differ from those used in \cite{Lin2019a} for PGT$^+$ by having
one additional row/column in both the $0^+$ and $1^-$ sectors, which
are related to the extra vector gauge field $B_A$ present in
WGT$^+$. For more details about SPOs in general, please refer to
\cite{Lin2019a}.

\section{No-tachyon and no-ghost conditions for the $1^-$ sector  \label{sec:MassiveCond1nWGT}}

First, to avoid tachyons and a dipole ghost, one requires the roots of
\eqref{eqn:WGTdet1n} to be real and distinct, such that
\begin{align*}
&\left\{6 c_1 t_1 \left(t_3-\lambda \right)+\left(r_1+r_4+r_5\right) \left[12 \left(t_3-\lambda \right) \left(t_1+\lambda \right)+\left(t_1+t_3\right) \nu \right]+6 t_1 t_3 \xi \right\}{}^2\\
&\qquad+3 t_1 \left(t_1+t_3\right) \left[12 \left(t_3-\lambda \right) \lambda +t_3 \nu \right] \left[c_1^2-8 \left(r_1+r_4+r_5\right) \xi \right]>0. \numberthis
\end{align*}
The no-tachyons conditions that both of the roots are positive then read
\begin{align*}
&\left(t_1+t_3\right) \left[c_1^2-8 \left(r_1+r_4+r_5\right) \xi \right] \left\{6 c_1 t_1 \left(t_3-\lambda \right)+\left(r_1+r_4+r_5\right) \left[12 \left(t_3-\lambda \right) \left(t_1+\lambda \right)+\left(t_1+t_3\right) \nu \right]+6 t_1 t_3 \xi \right\}>0,\numberthis\\
&t_1 \left(t_1+t_3\right) \left(12 \left(t_3-\lambda \right) \lambda +t_3 \nu \right) \left(c_1^2-8 \left(r_1+r_4+r_5\right) \xi \right)<0. \numberthis
\end{align*}

The no-ghost condition is
\begin{align*}
&\left[c_1^2-8 \left(r_1+r_4+r_5\right) \xi \right] \left[3 c_1 \left(t_1-2 t_3\right) \left(t_3-\lambda \right)-r_5 \left(t_1^2+2 t_1 t_3+19 t_3^2-36 t_3 \lambda +18 \lambda ^2\right)\right.\\&\qquad\left.-r_1 \left(t_1^2+2 t_1 t_3+19 t_3^2-36 t_3 \lambda +18 \lambda ^2\right)-r_4 \left(t_1^2+2 t_1 t_3+19 t_3^2-36 t_3 \lambda +18 \lambda ^2\right)-3 \left(t_1^2+2 t_3^2\right) \xi \right]>0, \numberthis\\
&\left(t_1+t_3\right)\left[c_1^2-8 \left(r_1+r_4+r_5\right) \xi \right] \left\{9 \left(t_1+t_3\right) \left\{2 t_1 \left(7 t_3^2-12 t_3 \lambda +6 \lambda ^2\right)+t_1^2 \left(14 t_3-12 \lambda +\nu \right)+2 t_3 \left[12 \left(t_3-\lambda \right) \lambda\right.\right.\right.\\ &\qquad\left.\left.\left.+t_3 \nu \right]\right\}{}^2 \left[c_1^2-8 \left(r_1+r_4+r_5\right) \xi \right]-48 t_1 \left[12 \left(t_3-\lambda \right) \lambda +t_3 \nu \right] \left[r_5 t_1^2-3 c_1 t_1 t_3+2 r_5 t_1 t_3+6 c_1 t_3^2+19 r_5 t_3^2\right.\right.\\&\qquad\left.\left.+3 c_1 t_1 \lambda -6 c_1 t_3 \lambda -36 r_5 t_3 \lambda +18 r_5 \lambda ^2+r_1 \left(t_1^2+2 t_1 t_3+19 t_3^2-36 t_3 \lambda +18 \lambda ^2\right)+r_4 \left(t_1^2+2 t_1 t_3+19 t_3^2\right.\right.\right.\\&\qquad\left.\left.\left.-36 t_3 \lambda +18 \lambda ^2\right)+3 t_1^2 \xi +6 t_3^2 \xi \right]{}^2+16 \left\{2 t_1 \left(7 t_3^2-12 t_3 \lambda +6 \lambda ^2\right)+t_1^2 \left(14 t_3-12 \lambda +\nu \right)+2 t_3 \left[12 \left(t_3-\lambda \right) \lambda +t_3 \nu \right]\right\}\right.\\ &\qquad\left.\left\{9 c_1 t_1 \left(-t_3+\lambda \right)+\frac{3}{2} \left(r_1+r_4+r_5\right) \left[-12 \left(t_3-\lambda \right) \left(t_1+\lambda \right)-\left(t_1+t_3\right) \nu \right]-9 t_1 t_3 \xi \right\} \left[-r_5 t_1^2+3 c_1 t_1 t_3-2 r_5 t_1 t_3\right.\right.\\&\qquad\left.\left.-6 c_1 t_3^2-19 r_5 t_3^2-3 c_1 t_1 \lambda +6 c_1 t_3 \lambda +36 r_5 t_3 \lambda -18 r_5 \lambda ^2-r_1 \left(t_1^2+2 t_1 t_3+19 t_3^2-36 t_3 \lambda +18 \lambda ^2\right)-r_4 \left(t_1^2+2 t_1 t_3\right.\right.\right.\\&\qquad\left.\left.\left.+19 t_3^2-36 t_3 \lambda +18 \lambda ^2\right)-3 \left(t_1^2+2 t_3^2\right) \xi \right]\right\}<0. \numberthis
\end{align*}

Combining the requirements for no tachyons and no ghosts, there exists
at least one parameter set satisfying all five conditions above, for
example
\begin{equation}
c_1= -5,r_1= 1,r_4= 0,r_5= 0,t_1= -4,t_3= -1,\lambda = -5,\nu = -244,\xi = 4,
\end{equation}
where the other parameters may take arbitrary values provided they do
not make the theory a critical case.  

\section{Completeness of the critical cases\label{sec:completeness}}

An ``additional condition'' is defined as the condition(s) to prevent
a theory from being critical. In our previous paper \cite{Lin2019a},
the additional condition was the requirement that the ``sibling
critical conditions'' should not be satisfied, and we will call
  this the ``sibling additional condition''. For example, consider a
theory that has the critical conditions that the (linear) parameter
combinations $X$, $Y$, and $Z$ should vanish; we will call $X$, $Y$
and $Z$ the ``critical parameters'' of the theory.  In the case, the
sibling critical parameters for the critical case $X=0$ are $Y$ and
$Z$.  To prevent a theory from being critical, one can require
  the ``critical parameters'' not equal to zeros. We will call this
  kind of condition a ``child additional condition''.  In PGT, as
discussed in \cite{Lin2019a}, the ``sibling additional condition'' is
identical to the ``child additional condition'', except for the root
case. This occurs because we add only one linear condition at a time
for cases resulting from type A or B critical conditions, but we
attempt to use all possible combinations of conditions simultaneously
for type C critical parameters (which we term
``combining'' the conditions).  We then recursively find the
  child critical cases of cases resulting from type A and B critical
  conditions  (the ``uncombined'' cases), but stop doing that for those
  from type C critical conditions (the ``combined'' cases).  If type
C critical conditions are treated in the same way as
type A and type B, then the statement is not valid for PGT.

There are two situations in which the statement is
  invalid. The first is the occurence of ``hidden'' critical
parameters. Consider a theory with only a $1\times 1$ $b$-matrix $(XY
+Z k^2)$. The theory has type B critical parameters, $X$ and $Y$, and
a type C one, $Z$. For the critical case $X=0$, the $b$-matrix becomes
($Z k^2$), so there is only one critical parameter $Z$. To prevent the
theory being critical (``child additional condition''), one requires
$Z\neq 0$. However, its sibling critical parameters are $Y$ and $Z$,
which are different. The critical parameter $Y$ is hidden in this
case. If there are ``hidden'' parameters and one is 
  requiring only child additional conditions, then a point in the
parameter space may belong to more than one critical case. For
example, the critical case $X=0,Z\neq 0$ and the case $Y=0, Z\neq 0$
has the overlap $X=Y=0,Z\neq0$, and they actually have the same
$b$-matrix $(Z k^2)$ and represent the same theory. If we use the
  sibling additional condition instead, the two cases become
  $X=0,Y\neq0,Z\neq0$ and $Y=0,X\neq 0,Z\neq 0$, and there is no
overlap.  ``Hidden'' parameters do not occur in PGT or any of the
critical cases discussed in this paper, if we ``combine'' all the type
C critical cases as in \cite{Lin2019a}. While the overlapping and
redundancy do no real harm to the correctness of our results, it may
be worth modifying our algorithm to accommodate the situation for
simplicity.

The second reason is the occurence of ``emergent'' critical
parameters. Some critical parameters appear after a $b$-matrix becomes
singular and a new $b$-matrix forms, which may happen in critical
cases resulting from a type A critical parameter (it is worth noting
that critical parameters of the root theory are always ``emergent''
because it has no parent or sibling critical cases). In PGT$^+$ and
torsion-free or simplified curvature-free WGT$^+$, either the new
$b$-matrix is $0\times 0$, or its critical parameters are already
included in the sibling critical parameters, and so there is no
``emergent'' critical parameter. However, in simplified full
WGT$^+$, this is not the case. For example, the $b(0^+)$-matrix of the
simplified root WGT$^+$ is
\begin{equation}
\left(
\begin{array}{ccc}
\makecell[cl]{2 \left[2 k^2 \left(r_1-r_3+2 r_4\right)+t_3\right]}& -2 i \sqrt{2} k t_3 & 2 \sqrt{6} \left(t_3-\lambda \right) \\
2 i \sqrt{2} k t_3 & 4 k^2 \left(t_3-\lambda \right) & 4 i \sqrt{3} k \left(t_3-\lambda \right) \\
2 \sqrt{6} \left(t_3-\lambda \right) & -4 i \sqrt{3} k \left(t_3-\lambda \right) & 12 \left(t_3-\lambda \right) \\
\end{array}
\right),
\end{equation}
which has $\det\left[b(0^+)\right]=-96 \left(t_3-\lambda \right) \lambda ^2  k^2$. Its critical case $\lambda=0$ has
\begin{equation}
\left(
\begin{array}{cc}
2 \left[2 k^2 \left(r_1-r_3+2 r_4\right)+t_3\right] & -2 i \sqrt{2} k t_3 \\
2 i \sqrt{2} k t_3 & 4 k^2 t_3 \\
\end{array}
\right)
\end{equation}
with $\det\left[b(0^+)\right]=16 \left(r_1-r_3+2 r_4\right) t_3
k^4$. The critical parameter $\left(r_1-r_3+2 r_4\right)$ is neither a
critical parameter of the root theory, nor among the sibling
critical parameters of case $\lambda=0$. However, the
``emergent'' parameters will not affect our algorithm if we apply the ``child additional condition'', which already includes the ``emergent'' parameters.

In conclusion, as long as there is no ``hidden'' critical parameter in
critical cases resulting from type A and B critical parameters, and
the cases resulting from type C critical parameters are ``combined'',
then we can apply the child additional conditions for the
``uncombined'' cases and the sibling additional conditions for the
``combined'' cases as the ``(extended) additional condition'',
respectively (this is also equivalent to combining the sibling and
child additional conditions as the additional condition for all
cases). This is what the term ``additional condition'' actually means
in this paper. Our algorithm then holds, and each parameter set
corresponds to one critical case. We have also checked that the all
critical cases in \cite{Lin2019a} and this paper cover the entire
parameter space and the critical cases have no overlap.

\section{Power-counting renormalizability\label{sec:PCRdef}}

Since the PCR criterion for PGT$^+$ is merely stated by Sezgin
\cite{Sezgin1980}, rather than derived, and we also wish to extend the
criterion to WGT$^+$, we give a brief outline derivation here.  Before
doing so, however, we note that PC is not the ultimate criterion for
renormalizability. Some PCR theories may be non-renormalizable because
of some deeper problems such as anomalies, and non-PCR theories may
turn out to be renormalizable (for example, see \cite{Parisi1975}).

We consider a quantum field theory in $d$ dimensional spacetime with
some fields labelled by $i$, and assume for each field the propagator
$\rightarrow k^{-l_i}$ as $k\rightarrow \infty$.  We also define the
canonical dimension \cite{Zinn-Justin2002} of the field $\varphi_i$ as
$[\varphi_i]\equiv (d-l_i)/2$, which only sometimes coincides with the
mass dimension of the field in natural units. The latter can be
inferred from the fact that each term in the Lagrangian density has
mass dimension $d$. One may always ensure that the two dimensions
coincide by making a field redefinition in which the original field is
multiplied by a constant. If the interactions are labelled by $a$,
with coupling constants $\lambda_a$, then the general criterion for a
theory to be PCR is that there is no coupling constant with negative
canonical dimension \cite{Zinn-Justin2002}, so that $[\lambda_a]\geq
0$ $\forall a$.

For WGT$^+$, in terms of the linearised fields introduced in \Cref{sec:root},
the most general Lagrangian in the Einstein gauge with
$\phi_0$ absorbed into the coefficients is given schematically by
\begin{align*}
b\mathcal{L}_G&\sim b\left(\lambda\mathcal{R}+r\mathcal{R}^2+t\mathcal{T}^{*2}+\xi \mathcal{H}^2+c_1\mathcal{R}\mathcal{H}+\nu B^2\right) \\
&\sim\left(1+f+f^2+...\right)\left\{\lambda\left(1+f\right)^2\left(\de A + A^2\right)+r\left(1+f\right)^4\left(\de A + A^2\right)^2\right.\\
&+t\left(1+f\right)^{2}\left[\de (f+f^2+...) + (1+f+f^2+...)(A+B)\right]^2+\xi(1+f)^4(\de B)^2\\
&\left.c_1(1+f+f^2+...)\left(\de A + A^2\right)\de B+\nu\left(1+f\right)^2B^2\right\},\numberthis
\end{align*}
where we do not show the detailed structures of the indices and
coefficients. The mass dimensions of the parameters and fields are
$[\lambda]_\text{M}=2$, $[r]_\text{M}=0$, $[t]_\text{M}=2$,
$[\xi]_\text{M}=0$, $[c_1]_\text{M}=0$, $[A]_\text{M}=1$,
$[f]_\text{M}=0$, and $[B]_\text{M}=1$.  Assuming the propagators of
$h$, $A$ and $B$ behave as $k^{-l_h}$, $k^{-l_A}$ and $k^{-l_B}$,
respectively, we need to redefine the fields as
$\tilde{h}=M_h^{2-l_h/2}h$, $\tilde{A}=M_A^{1-l_A/2}A$ and
$\tilde{B}=M_B^{2-l_B/2}B$. Therefore we require $l_h\geq 4$, $l_A\geq
2$ and $l_B\geq 2$ for the theory to be PCR.\footnote{If $r=0$, then
  the interaction terms with the highest degree of $A$ are $\sim A^2$
  with coefficients of dimension 2. Hence, in this case, we may have a
  looser condition $l_A\geq0$. However, there is no dynamical term for
  $A$ if $r=0$, so we consider $A$ not propagating.} The original PCR
criterion in \cite{Sezgin1980} for PGT$^+$ is obtained immediately by
setting $B=0$.

\section{PCR critical cases\label{sec:PCRcases}}

There exists a ``folk theorem'' dating back to the 1970s, a version of
which is presented in the introduction of Sezgin's paper
\cite{Sezgin1980}, that suggests that any gravity theory that is
unitary cannot also be PCR. The argument is not based on any rigorous
no-go theorem, but instead on the following simple observation: as shown in
\Cref{sec:PCRdef}, for a PGT$^+$ to be PCR the propagator of the $A$ field
must decay at least as quickly as $k^{-2}$ at high energy, and those
of the $\mathfrak{s}$ and $\mathfrak{a}$ fields must fall off at least
as $k^{-4}$, but the resulting total propagator, in general,
contains terms of opposite sign when expressed in partial fractions
and so the theory is not unitary. This viewpoint has never
subsequently been seriously challenged, and so our claim to have found
counterexamples is in conflict with the accepted wisdom. We therefore
take the opportunity here to elucidate the four unitary critical cases
that also satisfy the original criterion used by Sezgin in
\cite{Sezgin1980} to be PCR. These cases coincide with the PGT$^+$
cases 9, 10, 11 and 13, first identified in \cite{Lin2019a} and listed
in Table III of \cite{Lin2020}. In particular, we explain how these
theories, each of which contains only 2 massless d.o.f., evade the
argument in \cite{Sezgin1980}.

The key relevant property of these theories, at least in the
linearised approximation considered here, is that they contain no
`graviton' (d.o.f. associated with the $\mathfrak{s}$ and
$\mathfrak{a}$ fields), but only `tordions' (d.o.f. associated with
the $A$ field), as originally discussed in \cite{Lin2019a} (and no
`dilaton' d.o.f associated with the $B$ field, since we are
considering only PGT$^+$ here). In other words, for these four theories,
the $a$-matrices (\ref{eq:azerominus})--(\ref{eq:atwoplus}) contain
non-zero entries only in the rows/columns corresponding to the $A$
field. As a result, the propagator in each case need only decay at
least as quickly as $k^{-2}$ at high energy, and so the partial
fractions argument outlined above does not necessarily apply. 

One may verify directly by explicit calculation of their propagators
that this indeed occurs for cases 9, 10, 11 and 13. We consider each
case in turn, where the $a$-matrices for each case may be found by
substituting its critical condition into
(\ref{eq:azerominus})--(\ref{eq:atwoplus}).
\begin{enumerate}
\item For case 9, the critical condition is $r_2=r_1 -
  r_3 = r_4 = t_1 = t_2 = t_3 = \lambda = 0$, the resulting
  propagator of the $A$ field is
\begin{equation}
\hat{D}_A=\frac{1}{2\left(r_1+r_5\right)k^2}\hat{P}_{11}(1^-)+\frac{1}{2\left(2r_1+r_5\right)k^2}\hat{P}_{11}(1^+)+\frac{1}{2r_1 k^2}\hat{P}_{11}(2^-),
\end{equation}
and the condition for no ghost or tachyon is $r_1(r_1+r_5)(2r_1+r_5)<0$.
\item For case 10, the critical condition is $r_2 = r_1 = r_3/2 - r_4
  = t_1 = t_2 = t_3 = \lambda = 0$, the propagator is
\begin{equation}
\hat{D}_A=\frac{1}{\left(r_3+2r_5\right)k^2}\hat{P}_{11}(1^-)+\frac{1}{2\left(2r_3+r_5\right)k^2}\hat{P}_{11}(1^+)-\frac{1}{3r_3
  k^2}\hat{P}_{11}(2^+),
\end{equation}
and the condition for no ghost or tachyon is $r_3(2r_3+r_5)(r_3+ 2r_5)
<0$.
\item For case 11, the critical condition is $r_1 = r_3/2 - r_4 = t_1
  = t_2 = t_3 = \lambda = 0$, the propagator is
\begin{equation}
\hat{D}_A=\frac{1}{2r_2
  k^2}\hat{P}_{11}(0^-)+\frac{1}{\left(r_3+2r_5\right)k^2}\hat{P}_{11}(1^-)+\frac{1}{2\left(2r_3+r_5\right)k^2}\hat{P}_{11}(1^+)-\frac{1}{3r_3
  k^2}\hat{P}_{11}(2^+),
\end{equation}
and the condition for no ghost or tachyon is $r_3(2r_3+r_5) (r_3+ 2r_5)<0$.
\item For case 13, the critical condition is $r_2 = 2 r_1 - 2 r_3 +
  r_4 = t_1 = t_2 = t_3 = \lambda = 0$, the propagator is
\begin{equation}
\hat{D}_A=\frac{1}{-12\left(r_1-r_3\right)k^2}\hat{P}_{11}(0^+)+\frac{1}{2\left(-r_1+2r_3+r_5\right)k^2}\hat{P}_{11}(1^-)+\frac{1}{2\left(2r_3+r_5\right)k^2}\hat{P}_{11}(1^+)+\frac{1}{2r_1 k^2}\hat{P}_{11}(2^-),
\end{equation}
and the condition for no ghost or tachyon is
 $r_1 (r_1-2r_3-r_5) (2r_3+r_5)>0$.
\end{enumerate}
Since $\Theta_{AB}=\eta_{AB}-\frac{k_A k_B}{k^2}$ and $\Omega_{AB} =
\frac{k_A k_B}{k^2}$, all the SPOs behave as constants at high $k^2$.
Therefore, in each case the propagator of the $A$ field goes as
$k^{-2}$ at high energy and so the theory is PCR. We also note that,
for each case, the additional conditions that prevent the theory from
becoming a different critical case are that none of the demoninators
of the coefficients of the SPOs may vanish.

The absence of a `graviton' does not, however, preclude the
possibility that the 2 `tordion' massless d.o.f are in the spin $2^+$
sector, and indeed this may occur for cases 10 and 11, although not
for cases 9 and 13, as discussed in \cite{Lin2019a}; this is also
apparent from the above propagator for each theory. Thus, in cases 10
and 11, aspects of the gravitational interaction may still be mediated
by a massless spin-$2^+$ particle, despite it corresponding to
d.o.f. of the $A$ field rather than of the $\mathfrak{s}$ and
$\mathfrak{a}$ fields. As mentioned in \cite{Lin2019a}, it is worth
pointing out again here that the actions of cases 10 and 11 both
reduce in the absence of torsion to that of conformal gravity, which
is well known to be PCR but not unitary; it is claimed that one can nonetheless
construct a unitary quantum theory of conformal gravity by redefining
its Fock space \cite{Mannheim2011}, although this suggestion is
controversial \cite{Smilga2009}.

\twocolumngrid

\bibliography{GaugeGravity_abbr}

\end{document}
%